\input harvmac.tex

\input epsf.tex

\def\figin{\epsfcheck\figin}\def\figins{\epsfcheck\figins}
\def\epsfcheck{\ifx\epsfbox\UnDeFiNeD
\message{(NO epsf.tex, FIGURES WILL BE IGNORED)}
\gdef\figin##1{\vskip2in}\gdef\figins##1{\hskip.5in}% blank space instead
\else\message{(FIGURES WILL BE INCLUDED)}%
\gdef\figin##1{##1}\gdef\figins##1{##1}\fi}
\def\DefWarn#1{}
\def\figinsert{\goodbreak\midinsert}
\def\ifig#1#2#3{\DefWarn#1\xdef#1{figure~\the\figno}
\writedef{#1\leftbracket figure\noexpand~\the\figno}%
\figinsert\figin{\centerline{#3}}\medskip\centerline{\vbox{\baselineskip12pt
\advance\hsize by -1truein\noindent\footnotefont{\bf
Figure~\the\figno:} #2}}
\bigskip\endinsert\global\advance\figno by1}
%%% TO PUT FIGURES INSERT:
%\ifig\calorimeters{ A localized excitation is produced in a
%conformal field theory and its decay products are measured by
%calorimeters sitting far away.
%  } {\epsfxsize2.5in\epsfbox{calorimeters.eps}}

 % \draftmode

%%%%%%%%%%%%%%%%%%%%%%%%%%%%5

%\WittenXY
\lref\wittenbaryon{
  E.~Witten,
  ``Baryons and branes in anti de Sitter space,''
  JHEP {\bf 9807}, 006 (1998)
  [arXiv:hep-th/9805112].
  %%CITATION = JHEPA,9807,006;%%
}

\lref\kao{
  B.~M.~Zupnik and D.~V.~Khetselius,
  ``Three-dimensional extended supersymmetry in harmonic superspace,''
  Sov.\ J.\ Nucl.\ Phys.\  {\bf 47}, 730 (1988)
  [Yad.\ Fiz.\  {\bf 47}, 1147 (1988)];
  %%CITATION = YAFIA,47,1147;%%
  H.~C.~Kao,
  ``Selfdual Yang-Mills Chern-Simons Higgs systems with an N=3 extended
  supersymmetry,''
  Phys.\ Rev.\  D {\bf 50}, 2881 (1994);
  %%CITATION = PHRVA,D50,2881;%%
  }
 \lref\cscoefficient{ H.~C.~Kao, K.~M.~Lee and T.~Lee,
  ``The Chern-Simons coefficient in supersymmetric Yang-Mills Chern-Simons
  theories,''
  Phys.\ Lett.\  B {\bf 373}, 94 (1996)
  [arXiv:hep-th/9506170].
  %%CITATION = PHLTA,B373,94;%%
}

%\KlebanovUN
\lref\klebanovtseytlin{
  I.~R.~Klebanov and A.~A.~Tseytlin,
  ``Entropy of Near-Extremal Black p-branes,''
  Nucl.\ Phys.\  B {\bf 475}, 164 (1996)
  [arXiv:hep-th/9604089].
  %%CITATION = NUPHA,B475,164;%%
}

%\MaldacenaIM
\lref\MaldacenaIM{
  J.~M.~Maldacena,
  ``Wilson  loops in large N field theories,''
  Phys.\ Rev.\ Lett.\  {\bf 80}, 4859 (1998)
  [arXiv:hep-th/9803002];
  %%CITATION = PRLTA,80,4859;%%
  S.~J.~Rey and J.~T.~Yee,
  ``Macroscopic strings as heavy quarks in large N gauge theory and  anti-de
  Sitter supergravity,''
  Eur.\ Phys.\ J.\  C {\bf 22}, 379 (2001)
  [arXiv:hep-th/9803001].
  %%CITATION = EPHJA,C22,379;%%
}

%\MaldacenaRE
\lref\MaldacenaRE{
  J.~M.~Maldacena,
  ``The large N limit of superconformal field theories and supergravity,''
  Adv.\ Theor.\ Math.\ Phys.\  {\bf 2}, 231 (1998)
  [Int.\ J.\ Theor.\ Phys.\  {\bf 38}, 1113 (1999)]
  [arXiv:hep-th/9711200].
  %%CITATION = IJTPB,38,1113;%%
}

%\GoldschmidtWQ
\lref\wittencp{
  Y.~Y.~Goldschmidt and E.~Witten,
  ``Conservation Laws In Some Two-Dimensional Models,''
  Phys.\ Lett.\  B {\bf 91}, 392 (1980).
  %%CITATION = PHLTA,B91,392;%%
}

%\GomesQH
\lref\abdalla{
  M.~Gomes, E.~Abdalla and M.~C.~B.~Abdalla,
  ``On The Nonlocal Charge Of The $CP^{N-1}$ Model And Its Supersymmetric
  Extension To All Orders,''
  Phys.\ Rev.\  D {\bf 27}, 825 (1983).
  %%CITATION = PHRVA,D27,825;%%
}

 %%%%%%%%%%%%%%%%%%%%%%%%%%%%%%%%%%%%%%%%%%%%%%%5
%\GaiottoQI
\lref\gaiottoyin{
  D.~Gaiotto and X.~Yin,
  ``Notes on superconformal Chern-Simons-matter theories,''
  JHEP {\bf 0708}, 056 (2007)
  [arXiv:0704.3740 [hep-th]].
  %%CITATION = JHEPA,0708,056;%%
}

\lref\wittenthree{
  E.~Witten,
  ``Supersymmetric index of three-dimensional gauge theory,''
  arXiv:hep-th/9903005.
  %%CITATION = HEP-TH/9903005;%%
}

%\BiranIY
\lref\BiranIY{
  B.~Biran, A.~Casher, F.~Englert, M.~Rooman and P.~Spindel,
  ``The Fluctuating Seven Sphere In Eleven-Dimensional Supergravity,''
  Phys.\ Lett.\  B {\bf 134}, 179 (1984);
  %%CITATION = PHLTA,B134,179;%%
  L.~Castellani, R.~D'Auria, P.~Fre, K.~Pilch and P.~van Nieuwenhuizen,
  ``The Bosonic Mass Formula For Freund-Rubin Solutions Of D = 11 Supergravity
  On General Coset Manifolds,''
  Class.\ Quant.\ Grav.\  {\bf 1}, 339 (1984).
  %%CITATION = CQGRD,1,339;%%
}

%\SenZQ
\lref\senteight{
  A.~Sen,
  ``Orbifolds of M-Theory and String Theory,''
  Mod.\ Phys.\ Lett.\  A {\bf 11}, 1339 (1996)
  [arXiv:hep-th/9603113].
  %%CITATION = MPLAE,A11,1339;%%
}

%\GauntlettPK
\lref\kkmonopole{
  J.~P.~Gauntlett, G.~W.~Gibbons, G.~Papadopoulos and P.~K.~Townsend,
  ``Hyper-Kaehler manifolds and multiply intersecting branes,''
  Nucl.\ Phys.\  B {\bf 500}, 133 (1997)
  [arXiv:hep-th/9702202].
  %%CITATION = NUPHA,B500,133;%%
}

%\GoddardQE
\lref\GNO{
  P.~Goddard, J.~Nuyts and D.~I.~Olive,
  ``Gauge Theories And Magnetic Charge,''
  Nucl.\ Phys.\  B {\bf 125}, 1 (1977).
  %%CITATION = NUPHA,B125,1;%%
}

%\McGreevyCW
\lref\McGreevyCW{
  J.~McGreevy, L.~Susskind and N.~Toumbas,
  ``Invasion of the giant gravitons from anti-de Sitter space,''
  JHEP {\bf 0006}, 008 (2000)
  [arXiv:hep-th/0003075].
  %%CITATION = JHEPA,0006,008;%%
}

%\'tHooftHY
\lref\tHooftHY{
  G.~'t Hooft,
  ``On The Phase Transition Towards Permanent Quark Confinement,''
  Nucl.\ Phys.\  B {\bf 138}, 1 (1978).
  %%CITATION = NUPHA,B138,1;%%
}

%\AharonyRM
\lref\AharonyRM{
  O.~Aharony, Y.~Oz and Z.~Yin,
  ``M-theory on AdS(p) x S(11-p) and superconformal field theories,''
  Phys.\ Lett.\  B {\bf 430}, 87 (1998)
  [arXiv:hep-th/9803051];
  %%CITATION = PHLTA,B430,87;%%
  S.~Minwalla,
  ``Particles on AdS(4/7) and primary operators on M(2/5) brane
  worldvolumes,''
  JHEP {\bf 9810}, 002 (1998)
  [arXiv:hep-th/9803053];
  %%CITATION = JHEPA,9810,002;%%
  E.~Halyo,
  ``Supergravity on AdS(4/7) x S(7/4) and M branes,''
  JHEP {\bf 9804}, 011 (1998)
  [arXiv:hep-th/9803077].
  %%CITATION = JHEPA,9804,011;%%
}

%\ElitzurNR
\lref\emss{
  S.~Elitzur, G.~W.~Moore, A.~Schwimmer and N.~Seiberg,
  ``Remarks On The Canonical Quantization Of The Chern-Simons-Witten Theory,''
  Nucl.\ Phys.\  B {\bf 326}, 108 (1989).
  %%CITATION = NUPHA,B326,108;%%
}
%\BergmanNA
\lref\bergman{
  O.~Bergman, A.~Hanany, A.~Karch and B.~Kol,
  ``Branes and supersymmetry breaking in 3D gauge theories,''
  JHEP {\bf 9910}, 036 (1999)
  [arXiv:hep-th/9908075].
  %%CITATION = JHEPA,9910,036;%%
}
%\WittenYA
\lref\wittensl{
  E.~Witten,
  ``SL(2,Z) action on three-dimensional conformal field theories with Abelian
  symmetry,''
  arXiv:hep-th/0307041.
  %%CITATION = HEP-TH/0307041;%%
}
 %\TomasielloEQ
\lref\TomasielloEQ{
  A.~Tomasiello,
  ``New string vacua from twistor spaces,''
  arXiv:0712.1396 [hep-th].
  %%CITATION = ARXIV:0712.1396;%%
}
%\cite{Witten:1988hf}
\lref\wittenjones{
  E.~Witten,
  ``Quantum field theory and the Jones polynomial,''
  Commun.\ Math.\ Phys.\  {\bf 121}, 351 (1989).
  %%CITATION = CMPHA,121,351;%%
}

%\BeckerGJ
\lref\becker{
  K.~Becker and M.~Becker,
  ``M-Theory on Eight-Manifolds,''
  Nucl.\ Phys.\  B {\bf 477}, 155 (1996)
  [arXiv:hep-th/9605053].
  %%CITATION = NUPHA,B477,155;%%
}

%\LambertET
\lref\lamberttong{
  N.~Lambert and D.~Tong,
  ``Membranes on an Orbifold,''
  arXiv:0804.1114 [hep-th].
  %%CITATION = ARXIV:0804.1114;%%
}

%\HalyoPN
\lref\HalyoPN{
  E.~Halyo,
  ``Supergravity on AdS(5/4) x Hopf fibrations and conformal field  theories,''
  Mod.\ Phys.\ Lett.\  A {\bf 15}, 397 (2000)
  [arXiv:hep-th/9803193].
  %%CITATION = MPLAE,A15,397;%%
}

%\SethiZK
\lref\SethiZK{
  S.~Sethi,
  ``A relation between N = 8 gauge theories in three dimensions,''
  JHEP {\bf 9811}, 003 (1998)
  [arXiv:hep-th/9809162].
  %%CITATION = JHEPA,9811,003;%%
}

%\KapustinPY
\lref\kapustin{
  A.~Kapustin,
  ``Wilson-'t Hooft operators in four-dimensional gauge theories and
  S-duality,''
  Phys.\ Rev.\  D {\bf 74}, 025005 (2006)
  [arXiv:hep-th/0501015].
  %%CITATION = PHRVA,D74,025005;%%
  }

%\MartelliVH
\lref\martelli{
  D.~Martelli and J.~Sparks,
  ``Dual giant gravitons in Sasaki-Einstein backgrounds,''
  Nucl.\ Phys.\  B {\bf 759}, 292 (2006)
  [arXiv:hep-th/0608060].
  %%CITATION = NUPHA,B759,292;%%
}
 %\BorokhovIB
\lref\kapustinthooftone{
  V.~Borokhov, A.~Kapustin and X.~k.~Wu,
  ``Topological disorder operators in three-dimensional conformal field
  theory,''
  JHEP {\bf 0211}, 049 (2002)
  [arXiv:hep-th/0206054].
  %%CITATION = JHEPA,0211,049;%%
}

%\BorokhovCG
\lref\kapustinthoofttwo{
  V.~Borokhov, A.~Kapustin and X.~k.~Wu,
  ``Monopole operators and mirror symmetry in three dimensions,''
  JHEP {\bf 0212}, 044 (2002)
  [arXiv:hep-th/0207074].
  %%CITATION = JHEPA,0212,044;%%
}

%\BorokhovYU
\lref\kapustinthooftthree{
  V.~Borokhov,
  ``Monopole operators in three-dimensional N = 4 SYM and mirror symmetry,''
  JHEP {\bf 0403}, 008 (2004)
  [arXiv:hep-th/0310254].
  %%CITATION = JHEPA,0403,008;%%
}

%\DuffQZ
\lref\susywosusy{
  M.~J.~Duff, H.~Lu and C.~N.~Pope,
  ``Supersymmetry without supersymmetry,''
  Phys.\ Lett.\  B {\bf 409}, 136 (1997)
  [arXiv:hep-th/9704186].
  %%CITATION = PHLTA,B409,136;%%
}

  %\GubserTV
\lref\gkp{
  S.~S.~Gubser, I.~R.~Klebanov and A.~M.~Polyakov,
  ``A semi-classical limit of the gauge/string correspondence,''
  Nucl.\ Phys.\  B {\bf 636}, 99 (2002)
  [arXiv:hep-th/0204051].
  %%CITATION = NUPHA,B636,99;%%
}

%\AldayMF
\lref\aldayjmtwo{
  L.~F.~Alday and J.~M.~Maldacena,
  ``Comments on operators with large spin,''
  JHEP {\bf 0711}, 019 (2007)
  [arXiv:0708.0672 [hep-th]].
  %%CITATION = JHEPA,0711,019;%%
}

%\WittenZW
\lref\WittenZW{
  E.~Witten,
  ``Anti-de Sitter space, thermal phase transition, and confinement in  gauge
  theories,''
  Adv.\ Theor.\ Math.\ Phys.\  {\bf 2}, 505 (1998)
  [arXiv:hep-th/9803131].
  %%CITATION = 00203,2,505;%%
}

%\'tHooftJZ
\lref\tHooftJZ{
  G.~'t Hooft,
  ``A planar diagram theory for strong interactions,''
  Nucl.\ Phys.\  B {\bf 72}, 461 (1974).
  %%CITATION = NUPHA,B72,461;%%
}

%%%%%%%%%%%%% refernces added for introduction %%%%%%%%%%%%%%%%%

%\SchwarzYJ
\lref\schwarz{
  J.~H.~Schwarz,
  ``Superconformal Chern-Simons theories,''
  JHEP {\bf 0411}, 078 (2004)
  [arXiv:hep-th/0411077].
  %%CITATION = JHEPA,0411,078;%%
}

%\GubserDE
\lref\GubserDE{
  S.~S.~Gubser, I.~R.~Klebanov and A.~W.~Peet,
  ``Entropy and Temperature of Black 3-Branes,''
  Phys.\ Rev.\  D {\bf 54}, 3915 (1996)
  [arXiv:hep-th/9602135].
  %%CITATION = PHRVA,D54,3915;%%
}

%\BaggerVI
\lref\baggerlambert{
  J.~Bagger and N.~Lambert,
  ``Comments On Multiple M2-branes,''
  JHEP {\bf 0802}, 105 (2008)
  [arXiv:0712.3738 [hep-th]];
  %%CITATION = JHEPA,0802,105;%%
  J.~Bagger and N.~Lambert,
  ``Gauge Symmetry and Supersymmetry of Multiple M2-Branes,''
  Phys.\ Rev.\  D {\bf 77}, 065008 (2008)
  [arXiv:0711.0955 [hep-th]];
  %%CITATION = PHRVA,D77,065008;%%
  J.~Bagger and N.~Lambert,
  ``Modeling multiple M2's,''
  Phys.\ Rev.\  D {\bf 75}, 045020 (2007)
  [arXiv:hep-th/0611108].
  %%CITATION = PHRVA,D75,045020;%%
}

%\VanRaamsdonkFT
\lref\markvr{
  M.~Van Raamsdonk,
  ``Comments on the Bagger-Lambert theory and multiple M2-branes,''
  arXiv:0803.3803 [hep-th].
  %%CITATION = ARXIV:0803.3803;%%
}

%%%%%%%%% References added from section2.tex %%%%%%%%%%%%%%%%%%%%%%%%%

%\KlebanovHH
\lref\KlebanovHH{
  I.~R.~Klebanov and E.~Witten,
  ``Superconformal field theory on threebranes at a Calabi-Yau  singularity,''
  Nucl.\ Phys.\  B {\bf 536}, 199 (1998)
  [arXiv:hep-th/9807080].
  %%CITATION = NUPHA,B536,199;%%
}

%\DistlerMK
\lref\dmpv{
  J.~Distler, S.~Mukhi, C.~Papageorgakis and M.~Van Raamsdonk,
  ``M2-branes on M-folds,''
  JHEP {\bf 0805}, 038 (2008)
  [arXiv:0804.1256 [hep-th]].
  %%CITATION = JHEPA,0805,038;%%
}

%\ItzhakiRC
\lref\ItzhakiRC{
  N.~Itzhaki,
  ``Anyons, 't Hooft loops and a generalized connection in three dimensions,''
  Phys.\ Rev.\  D {\bf 67}, 065008 (2003)
  [arXiv:hep-th/0211140].
  %%CITATION = PHRVA,D67,065008;%%
}

%\LambertET
\lref\lamberttong{
  N.~Lambert and D.~Tong,
  ``Membranes on an Orbifold,''
  arXiv:0804.1114 [hep-th].
  %%CITATION = ARXIV:0804.1114;%%
}

%\IntriligatorEX
\lref\IntriligatorEX{
  K.~A.~Intriligator and N.~Seiberg,
  ``Mirror symmetry in three dimensional gauge theories,''
  Phys.\ Lett.\  B {\bf 387}, 513 (1996)
  [arXiv:hep-th/9607207].
  %%CITATION = PHLTA,B387,513;%%
}

%\BaggerVI
\lref\baggerlambert{
  J.~Bagger and N.~Lambert,
  ``Comments On Multiple M2-branes,''
  JHEP {\bf 0802}, 105 (2008)
  [arXiv:0712.3738 [hep-th]];
  %%CITATION = JHEPA,0802,105;%%
  J.~Bagger and N.~Lambert,
  ``Gauge Symmetry and Supersymmetry of Multiple M2-Branes,''
  Phys.\ Rev.\  D {\bf 77}, 065008 (2008)
  [arXiv:0711.0955 [hep-th]];
  %%CITATION = PHRVA,D77,065008;%%
  J.~Bagger and N.~Lambert,
  ``Modeling multiple M2's,''
  Phys.\ Rev.\  D {\bf 75}, 045020 (2007)
  [arXiv:hep-th/0611108].
  %%CITATION = PHRVA,D75,045020;%%
}

\lref\nonrenold{
  A.~N.~Kapustin and P.~I.~Pronin,
  ``Nonrenormalization theorem for gauge coupling in (2+1)-dimensions,''
  Mod.\ Phys.\ Lett.\  A {\bf 9}, 1925 (1994)
  [arXiv:hep-th/9401053].
  %%CITATION = MPLAE,A9,1925;%%
 W.~Chen, G.~W.~Semenoff and Y.~S.~Wu,
  ``Two loop analysis of nonAbelian Chern-Simons theory,''
  Phys.\ Rev.\  D {\bf 46}, 5521 (1992)
  [arXiv:hep-th/9209005].
  %%CITATION = PHRVA,D46,5521;%%
  }
  \lref\kapustinstrassler{
    A.~Kapustin and M.~J.~Strassler,
  ``On mirror symmetry in three dimensional Abelian gauge theories,''
  JHEP {\bf 9904}, 021 (1999)
  [arXiv:hep-th/9902033].
  %%CITATION = JHEPA,9904,021;%%
}

\lref\GustavssonVU{
  A.~Gustavsson,
  ``Algebraic structures on parallel M2-branes,''
  arXiv:0709.1260 [hep-th];
  %%CITATION = ARXIV:0709.1260;%%
    A.~Gustavsson,
  ``Selfdual strings and loop space Nahm equations,''
  JHEP {\bf 0804}, 083 (2008)
  [arXiv:0802.3456 [hep-th]].
  %%CITATION = JHEPA,0804,083;%%
}

%\DuffQZ
\lref\susywosusy{
  M.~J.~Duff, H.~Lu and C.~N.~Pope,
  ``Supersymmetry without supersymmetry,''
  Phys.\ Lett.\  B {\bf 409}, 136 (1997)
  [arXiv:hep-th/9704186].
  %%CITATION = PHLTA,B409,136;%%
}
%% Refer to this paper in the sugra section.

%\MukhiUX
\lref\dtwomtwo{
  S.~Mukhi and C.~Papageorgakis,
  ``M2 to D2,''
  arXiv:0803.3218 [hep-th].
  %%CITATION = ARXIV:0803.3218;%%
}

%\ArkaniHamedIE
\lref\nimadec{
  N.~Arkani-Hamed, A.~G.~Cohen, D.~B.~Kaplan, A.~Karch and L.~Motl,
  ``Deconstructing (2,0) and little string theories,''
  JHEP {\bf 0301}, 083 (2003)
  [arXiv:hep-th/0110146].
  %%CITATION = JHEPA,0301,083;%%
}
\lref\mooreseiberg{
  G.~W.~Moore and N.~Seiberg,
  ``Taming the Conformal Zoo,''
  Phys.\ Lett.\  B {\bf 220}, 422 (1989).
  %%CITATION = PHLTA,B220,422;%%
  }
%\cite{Moore:1988qv}

%\ZupnikEN
\lref\csntworefs{
  B.~M.~Zupnik and D.~G.~Pak,
  ``Superfield formulation of the simplest three-dimensional gauge theories and
  conformal supergravities,''
  Theor.\ Math.\ Phys.\  {\bf 77}, 1070 (1988)
  [Teor.\ Mat.\ Fiz.\  {\bf 77}, 97 (1988)];
  %%CITATION = TMFZA,77,97;%%
% OB Added Lee, Lee and Weinberg reference here
%\LeeIT
  C.~k.~Lee, K.~M.~Lee and E.~J.~Weinberg,
  ``Supersymmetry And Selfdual Chern-Simons Systems,''
  Phys.\ Lett.\  B {\bf 243}, 105 (1990);
  %%CITATION = PHLTA,B243,105;%%
  E.~A.~Ivanov,
  ``Chern-Simons matter systems with manifest N=2 supersymmetry,''
  Phys.\ Lett.\  B {\bf 268}, 203 (1991);
  %%CITATION = PHLTA,B268,203;%%
  L.~V.~Avdeev, G.~V.~Grigorev and D.~I.~Kazakov,
  ``Renormalizations in Abelian Chern-Simons field theories with matter,''
  Nucl.\ Phys.\  B {\bf 382}, 561 (1992);
  %%CITATION = NUPHA,B382,561;%%
  L.~V.~Avdeev, D.~I.~Kazakov and I.~N.~Kondrashuk,
  ``Renormalizations in supersymmetric and nonsupersymmetric nonAbelian
  Chern-Simons field theories with matter,''
  Nucl.\ Phys.\  B {\bf 391}, 333 (1993);
  %%CITATION = NUPHA,B391,333;%%
  % JM Added the Gates references here
  S.~J.~J.~Gates and H.~Nishino,
  ``Remarks on the N=2 supersymmetric Chern-Simons theories,''
  Phys.\ Lett.\  B {\bf 281}, 72 (1992);
  %%CITATION = PHLTA,B281,72;%%
  H.~Nishino and S.~J.~J.~Gates,
  ``Chern-Simons theories with supersymmetries in three-dimensions,''
  Int.\ J.\ Mod.\ Phys.\  A {\bf 8}, 3371 (1993);
  %%CITATION = IMPAE,A8,3371;%%
  R.~Brooks and S.~J.~J.~Gates,
  ``Extended supersymmetry and superBF gauge theories,''
  Nucl.\ Phys.\  B {\bf 432}, 205 (1994)
  [arXiv:hep-th/9407147].
  %%CITATION = NUPHA,B432,205;%%
}

%\NilssonBJ
\lref\nilssonpope{
  B.~E.~W.~Nilsson and C.~N.~Pope,
  ``Hopf Fibration Of Eleven-Dimensional Supergravity,''
  Class.\ Quant.\ Grav.\  {\bf 1}, 499 (1984).
  %%CITATION = CQGRD,1,499;%%
}

\lref\WatamuraHJ{
  S.~Watamura,
  ``Spontaneous Compactification And $ CP^N: ~SU(3)\times  SU(2) \times U(1)$,
  $\sin^2\theta_W$, G(3) / G(2) And SU(3) Triplet Chiral Fermions In
  Four-Dimensions,''
  Phys.\ Lett.\  B {\bf 136}, 245 (1984).
  %%CITATION = PHLTA,B136,245;%%
}
\lref\HosomichiJD{
  K.~Hosomichi, K.~M.~Lee, S.~Lee, S.~Lee and J.~Park,
  ``N=4 Superconformal Chern-Simons Theories with Hyper and Twisted Hyper
  Multiplets,''
  arXiv:0805.3662 [hep-th].
  %%CITATION = ARXIV:0805.3662;%%
}

\lref\GaiottoSD{
  D.~Gaiotto and E.~Witten,
  ``Janus Configurations, Chern-Simons Couplings, And The Theta-Angle in N=4
  Super Yang-Mills Theory,''
  arXiv:0804.2907 [hep-th].
  %%CITATION = ARXIV:0804.2907;%%
}

\lref\gibbonshawking{
  G.~W.~Gibbons and S.~W.~Hawking,
  ``Gravitational Multi - Instantons,''
  Phys.\ Lett.\  B {\bf 78}, 430 (1978).
  %%CITATION = PHLTA,B78,430;%%
  }

\lref\hananybaryon{
  D.~Forcella, A.~Hanany, Y.~H.~He and A.~Zaffaroni,
  ``The Master Space of N=1 Gauge Theories,''
  arXiv:0801.1585 [hep-th].
  %%CITATION = ARXIV:0801.1585;%%
}

%\FengUR
\lref\hananyexp{
  B.~Feng, A.~Hanany and Y.~H.~He,
  ``Counting gauge invariants: The plethystic program,''
  JHEP {\bf 0703}, 090 (2007)
  [arXiv:hep-th/0701063].
  %%CITATION = JHEPA,0703,090;%%
}

\lref\indexref{
  J.~Kinney, J.~M.~Maldacena, S.~Minwalla and S.~Raju,
  ``An index for 4 dimensional super conformal theories,''
  Commun.\ Math.\ Phys.\  {\bf 275}, 209 (2007)
  [arXiv:hep-th/0510251].
  %%CITATION = CMPHA,275,209;%%
}

%\MaldacenaSS
\lref\MaldacenaSS{
  J.~M.~Maldacena, G.~W.~Moore and N.~Seiberg,
  ``D-brane charges in five-brane backgrounds,''
  JHEP {\bf 0110}, 005 (2001)
  [arXiv:hep-th/0108152].
  %%CITATION = JHEPA,0110,005;%%
}

\lref\AharonyQU{
  O.~Aharony and E.~Witten,
  ``Anti-de Sitter space and the center of the gauge group,''
  JHEP {\bf 9811}, 018 (1998)
  [arXiv:hep-th/9807205].
  %%CITATION = JHEPA,9811,018;%%
}

%\GubserFP
\lref\GubserFP{
  S.~S.~Gubser and I.~R.~Klebanov,
  ``Baryons and domain walls in an N = 1 superconformal gauge theory,''
  Phys.\ Rev.\  D {\bf 58}, 125025 (1998)
  [arXiv:hep-th/9808075].
  %%CITATION = PHRVA,D58,125025;%%
}

%\BenaZB
\lref\BenaZB{
  I.~Bena,
  ``The M-theory dual of a 3 dimensional theory with reduced supersymmetry,''
  Phys.\ Rev.\  D {\bf 62}, 126006 (2000)
  [arXiv:hep-th/0004142].
  %%CITATION = PHRVA,D62,126006;%%
}

% Added references:

%\AharonyJU
\lref\AharonyJU{
  O.~Aharony and A.~Hanany,
  ``Branes, superpotentials and superconformal fixed points,''
  Nucl.\ Phys.\  B {\bf 504}, 239 (1997)
  [arXiv:hep-th/9704170].
  %%CITATION = NUPHA,B504,239;%%
}
%\cite{Niemi:1983rq}
\lref\niemimics{
 A.~J.~Niemi and G.~W.~Semenoff,
 ``Axial Anomaly Induced Fermion Fractionization And Effective Gauge Theory
 Actions In Odd Dimensional Space-Times,''
 Phys.\ Rev.\ Lett.\  {\bf 51}, 2077 (1983).
 %%CITATION = PRLTA,51,2077;%%
}
%\AlvarezGaumeIG
\lref\AlvarezGaumeIG{
  L.~Alvarez-Gaume and E.~Witten,
  ``Gravitational Anomalies,''
  Nucl.\ Phys.\  B {\bf 234}, 269 (1984).
  %%CITATION = NUPHA,B234,269;%%
}

%\SenZB
\lref\SenZB{
  A.~Sen,
  ``Kaluza-Klein dyons in string theory,''
  Phys.\ Rev.\ Lett.\  {\bf 79}, 1619 (1997)
  [arXiv:hep-th/9705212].
  %%CITATION = PRLTA,79,1619;%%
}
%\RedlichDV
\lref\RedlichDV{
  A.~N.~Redlich,
  ``Parity Violation And Gauge Noninvariance Of The Effective Gauge Field
  Action In Three-Dimensions,''
  Phys.\ Rev.\  D {\bf 29}, 2366 (1984).
  %%CITATION = PHRVA,D29,2366;%%
}
%\KitaoMF
\lref\KitaoMF{
  T.~Kitao, K.~Ohta and N.~Ohta,
  ``Three-dimensional gauge dynamics from brane configurations with
  (p,q)-fivebrane,''
  Nucl.\ Phys.\  B {\bf 539}, 79 (1999)
  [arXiv:hep-th/9808111].
  %%CITATION = NUPHA,B539,79;%%
}

%\HananyIE
\lref\HananyIE{
  A.~Hanany and E.~Witten,
  ``Type IIB superstrings, BPS monopoles, and three-dimensional gauge
  dynamics,''
  Nucl.\ Phys.\  B {\bf 492}, 152 (1997)
  [arXiv:hep-th/9611230].
  %%CITATION = NUPHA,B492,152;%%
}

%\GaiottoSD
\lref\GW{
  D.~Gaiotto and E.~Witten,
  ``Janus Configurations, Chern-Simons Couplings, And The Theta-Angle in N=4
  Super Yang-Mills Theory,''
  arXiv:0804.2907 [hep-th].
  %%CITATION = ARXIV:0804.2907;%%
}

\lref\multiplemtwo{
  J.~Gomis, G.~Milanesi and J.~G.~Russo,
  ``Bagger-Lambert Theory for General Lie Algebras,''
  arXiv:0805.1012 [hep-th];
  %%CITATION = ARXIV:0805.1012;%%
  S.~Benvenuti, D.~Rodriguez-Gomez, E.~Tonni and H.~Verlinde,
  ``N=8 superconformal gauge theories and M2 branes,''
  arXiv:0805.1087 [hep-th];
  %%CITATION = ARXIV:0805.1087;%%
  P.~M.~Ho, Y.~Imamura and Y.~Matsuo,
  ``M2 to D2 revisited,''
  arXiv:0805.1202 [hep-th];
  %%CITATION = ARXIV:0805.1202;%%
  M.~A.~Bandres, A.~E.~Lipstein and J.~H.~Schwarz,
  ``Ghost-Free Superconformal Action for Multiple M2-Branes,''
  arXiv:0806.0054 [hep-th];
  %%CITATION = ARXIV:0806.0054;%%
  J.~Gomis, D.~Rodriguez-Gomez, M.~Van Raamsdonk and H.~Verlinde,
  ``The Superconformal Gauge Theory on M2-Branes,''
  arXiv:0806.0738 [hep-th].
  %%CITATION = ARXIV:0806.0738;%%
}

%\LinNH
\lref\linjm{
  H.~Lin and J.~M.~Maldacena,
  ``Fivebranes from gauge theory,''
  Phys.\ Rev.\  D {\bf 74}, 084014 (2006)
  [arXiv:hep-th/0509235].
  %%CITATION = PHRVA,D74,084014;%%
}

%%%%%%%%%%%%%%%%%%%%%%%%%%%%%%%%%%5

%\draftmode

%=====================================================================

\Title{\vbox{\baselineskip12pt\hbox{}
\hbox{WIS/12/08-JUN-DPP}}} {\vbox{ \vskip -5cm
{\centerline{  ${\cal N}=6$ superconformal Chern-Simons-matter}
\medskip
\centerline{theories, M2-branes and their gravity duals}}}}

\vskip  -5mm
 \centerline{Ofer Aharony$^{a}$, Oren Bergman$^{b,c}$, Daniel Louis Jafferis$^{d}$ and Juan Maldacena$^{b}$}

\bigskip
{\sl
\centerline{$^a$Department of Particle Physics}
\centerline{The Weizmann Institute of Science, Rehovot 76100, Israel}
\centerline{\tt Ofer.Aharony@weizmann.ac.il}
\medskip
\centerline{$^{b}$School of Natural Sciences, Institute for Advanced Study, Princeton, NJ 08540, USA}
\centerline{\tt bergman@sns.ias.edu, malda@ias.edu}
\medskip
\centerline{$^{c}$Department of Physics, Technion, Haifa 32000, Israel}
\centerline{\tt bergman@physics.technion.ac.il}
\medskip
\centerline{$^{d}$Department of Physics, Rutgers University, Piscataway, NJ 08855, USA}
\centerline{\tt jafferis@physics.rutgers.edu}
\bigskip
\bigskip \medskip
}

% ABSTRACT

\leftskip 8mm  \rightskip 8mm \baselineskip14pt \noindent
%\tenrm
%{\tenbf \hskip 7mm Abstract.}
%
We construct three dimensional Chern-Simons-matter theories with gauge groups
$U(N)\times U(N)$ and $SU(N)\times SU(N)$ which have explicit ${\cal N}=6$ superconformal
symmetry.
Using brane constructions we   argue that the
 $U(N)\times U(N)$ theory at level $k$ describes the low energy limit of $N$ M2-branes probing a
 ${\bf C}^4/{\bf Z}_k$ singularity. At large $N$ the theory is then dual to
  M-theory on $AdS_4\times S^7/{\bf Z}_k$. The theory also has a 't Hooft limit (of large
  $N$ with a fixed ratio $N/k$) which is
  dual to type IIA string theory on $AdS_4 \times {\bf CP}^3$. For $k=1$ the theory
  is conjectured to describe $N$ M2-branes in flat space, although our construction realizes
  explicitly only
  six of the eight supersymmetries.  We
give some evidence for this conjecture, which is similar to the
evidence for mirror symmetry in $d=3$ gauge theories.
When the gauge group is $SU(2) \times SU(2)$ our theory has extra symmetries and
becomes identical to the Bagger-Lambert theory.

\bigskip\medskip

\leftskip 0mm  \rightskip 0mm
 \Date{\hskip 8mm June 2008}

\newsec{ Introduction}

\def\cn{ {\cal N} }

In this paper we  construct and study
conformal field theories in three dimensions with $\cn=6$ supersymmetry, or
a total of $12$ real supercharges.

Highly supersymmetric three dimensional conformal field theories
are interesting for various reasons. One is the construction of
the  theory  describing  the worldvolume of membranes in M-theory
(M2-branes) at low energies. Conformal Chern-Simons theories were
explored in \schwarz\ for this purpose, but these theories did not
have enough supersymmetry. More recently a theory with $\cn=8$
supersymmetry was constructed by Bagger and Lambert
\baggerlambert\ (see also \GustavssonVU) and was conjectured to be
related to a specific theory of M2-branes for $k=1,2$
\refs{\dmpv,\lamberttong}.

Another motivation  to study three dimensional conformal field
theories is that they could describe interesting conformal fixed
points in condensed matter systems. From this point of view the
highly supersymmetric versions are interesting toy models which
are more solvable. Of course, Chern-Simons theories often arise in
interesting condensed matter systems.

Finally, one is interested in examples of the $AdS_4/CFT_3$ correspondence as a way to study
some portion of the landscape of four dimensional backgrounds of string theory with a
negative cosmological constant (which admit $AdS_4$ solutions).
Here we find an infinite sequence of theories with a discrete coupling $k$, the level of the Chern-Simons
theory, such that for large $k$ the field theory has a weak coupling description while
for small $k$ the theory becomes strongly coupled. For the highly supersymmetric theories that
we consider, the size of
the internal space is comparable to the radius of curvature of $AdS_4$, so these backgrounds
do not really look four dimensional. However, this system might
be a good stepping stone in order to study less symmetric and more realistic compactifications.

The main theory we consider is a $d=3$, $U(N) \times U(N)$ gauge
theory with four complex scalar fields $C_I$ ($I=1,2,3,4$)
in the $\bf(N,\bar N)$ representation  and their corresponding complex conjugate fields in the
 $\bf(\bar N , N)$ representation. In addition, we have fermionic partners of these bosonic fields.
 The gauge fields are not dynamical, and have a
 Chern-Simons action with opposite integer levels for the
 two gauge groups, $k$ and
 $-k$. The theory we build has $\cn=6$ supersymmetry,
  and it is weakly coupled in the large $k$ limit ($k \gg N$).

 These theories can be obtained as the IR limit of a particular brane construction, similar
 to the ones considered in \refs{\KitaoMF,\bergman}.  The brane
 construction preserves only $\cn=3$ supersymmetry\foot{There is also another brane
 construction leading to the same IR theory, which preserves $\cn=4$ supersymmetry.}.
 The corresponding field theory has a matter content
 similar to what we had above, except that we now have dynamical gauge fields and their superpartners,
 which are
 massive due to the supersymmetric
 Chern-Simons terms. When we go to low energies we can integrate them out and recover the
 conformal  theory
 with only Chern-Simons terms.
  The low energy theory has an enhanced $\cn =6$ supersymmetry.

 The brane construction   can be lifted to M-theory where it
 corresponds to M2-branes probing a transverse toric HyperK\"ahler manifold \kkmonopole\ which
 describes
 two Kaluza-Klein monopoles at an  angle. This eight dimensional space
  has a singularity at a single point, which is
 of the form ${\bf R}^8/{\bf Z}_k$, or ${\bf C}^4/{\bf Z}_k$, where the ${\bf Z}_k$ acts
 by rotating the phases of all four
 complex coordinates. Thus, the conformal field theory that we are discussing is
 equivalent to the low-energy theory on
 $N$ M2-branes at this ${\bf C}^4/{\bf Z}_k$ singularity.

 Taking the large $N$ limit we can construct the gravity dual to these theories, which simply
 corresponds to M-theory on $AdS_4 \times S^7/{\bf Z}_k$. This description is weakly curved
 when $N \gg k^5$. For larger values of $k$ a circle in the M-theory
 description becomes small, and the more appropriate description is in terms
 of type IIA string theory on $AdS_4 \times {\bf CP}^3$.
 This string theory background also describes the 't Hooft limit of the theory,
 $N, k \to \infty$ with $\lambda = N/k$ fixed. The radius of
 curvature
 scales as $R^2 \sim \sqrt{\lambda } \alpha'$, so it is weakly curved when $k \ll N$.  We explore various properties of these
 solutions.

 From the field theory point of view one can also consider $\cn =6$ theories with an
 $SU(N) \times SU(N)$ (or $(SU(N)\times SU(N)) / {\bf Z}_N$)
 gauge group and the same structure as our theories. In the particular
 case of $N=2$ one finds some extra symmetries, due to the fact that the ${\bf 2}$ and ${\bf \bar 2}$
 representations of $SU(2)$ are equivalent. These extra symmetries imply that the
 corresponding Chern-Simons-matter theory has
 $\cn =8$ supersymmetry, and in fact it is precisely equivalent to the Bagger-Lambert theory
 \baggerlambert\ in the presentation given in \markvr. In our formulation of this theory
 the 3-algebra structure introduced in \refs{\baggerlambert,\GustavssonVU} does not play any role.

 In the case of $k=1$ our theories should reduce to the theory of $N$ M2-branes in flat space,
 and for $k=2$ they reduce to the theory of $N$ M2-branes on ${\bf R}^8/{\bf Z}_2$. In these
 cases the theory should have additional supersymmetries (${\cal N}=8$) that are not apparent in the Lagrangian, since they
 exist only for these specific values of the coupling. We have not found a direct proof of this statement from
 the field theory point of view, but it is clear from the brane construction or string theory arguments, and we provide various tests of this statement.
 A good analogy would be the theory of a compact
 boson in two dimensions, which is generically not supersymmetric but becomes supersymmetric at
 a special value of the radius of the circle
 %\GinspargUI
\ref\GinspargUI{
  P.~H.~Ginsparg,
  ``Applied Conformal Field Theory,''
  arXiv:hep-th/9108028.
  %%CITATION = HEP-TH/9108028;%%
}.
In any case, we claim that the sequence of theories that we
explore contains the $\cn =8$
  M2-brane theory
 as a particular example (whose field theory description is strongly coupled).

 This paper is organized as follows. In section two we construct the three
 dimensional theories we will be considering and analyze their properties. In section three we discuss the brane construction
 of the same theories, and argue that they reduce to M2-branes probing a ${\bf C}^4/{\bf Z}_k$ singularity.
 In section four we discuss aspects of the gravity dual of these theories.
We end in section five with some conclusions and open problems.
Two appendices provide some technical details.

\newsec{Chern-Simons-matter theories with ${\cal N} \ge 6$ superconformal symmetry}

In this section we construct and analyze Chern-Simons matter theories with
 gauge groups $U(N)\times U(N)$
and $SU(N)\times SU(N)$ which have an explicit ${\cal N}=6$ superconformal symmetry. We
begin in section 2.1 by reviewing the construction of Chern-Simons-matter theories with
${\cal N}=2$ and ${\cal N}=3$ supersymmetry, which exist for arbitrary gauge groups and
matter content. In section 2.2 we show that for specific gauge groups and matter
content, the supersymmetry is enhanced to ${\cal N}=6$. In section 2.3 we compute the moduli
space of the $U(N)\times U(N)$ theories at level $k$, and show that it is the same as that
of $N$ M2-branes probing a ${\bf C}^4/{\bf Z}_k$ singularity. In section 2.4 we compute the
spectrum of chiral operators and Wilson lines in these theories. In section 2.5 we discuss the
$SU(N)\times SU(N)$ case.
In section 2.6 we argue that for $N=2$ these theories have an enhanced ${\cal N}=8$
supersymmetry, and we compare with the Bagger-Lambert theories \baggerlambert .

\subsec{A review of ${\cal N}=2,3$ Chern-Simons-matter theories}

Chern-Simons theories in $2+1$ dimensions are simple examples of topological theories.
When they are coupled to matter fields, the theory is no longer topological; however,
in some cases it is still conformally invariant. A specific example which was argued
\refs{\csntworefs,\gaiottoyin}
to be exactly conformal is the case of ${\cal N}=2$ supersymmetric Chern-Simons-matter
theories with no superpotential. The field content of such theories includes a vector
multiplet $V$ (which is the dimensional reduction of the four dimensional ${\cal N}=1$
vector multiplet) in the adjoint of the gauge group $G$, and chiral multiplets $\Phi_i$
in representations $R_i$ of this group. The kinetic term for the chiral multiplets takes
the usual form (which is the dimensional reduction of the four dimensional kinetic term).
However, the kinetic term for the vector multiplet is replaced by a
supersymmetric Chern-Simons term. The form of this term in superspace is somewhat
awkward, but in components (in Wess-Zumino gauge) it takes the simple form\foot{We
will generally adhere to the notations and conventions of \gaiottoyin, except for some
factors of $2$ in the ${\cal N}=3$ superpotentials which seem to be essential for
obtaining theories with ${\cal N}=3$ supersymmetry.}
\eqn\csntwo{S_{CS}^{{\cal N}=2} = {k\over {4\pi}} \int \Tr(A \wedge dA +
{2\over 3} A^3 - {\bar \chi} \chi + 2 D \sigma),}
where $\chi$ is the gaugino, $D$ is the auxiliary field of the vector multiplet, and
$\sigma$ is the real scalar field in the vector multiplet (coming from the $A_3$
component of the gauge field when we dimensionally reduce from $3+1$ dimensions).
For non-Abelian theories the level $k$ is quantized; for $SU(N)$ or $U(N)$ theories it
is quantized to be an integer when the trace in \csntwo\ is in the fundamental representation.

Note that there is no kinetic term for any of the fields in the
vector multiplet, so they are all auxiliary fields. The kinetic
term for the chiral multiplets includes couplings $-\bar\phi_i
\sigma^2 \phi_i - \bar{\psi}_i \sigma \psi_i$, which are the
dimensional reduction of the kinetic terms in the fourth
direction. In addition we have the usual $D$ term coupling $\bar
\phi_i D \phi_i$.
 We can integrate out the $D$ field and obtain
$\sigma = -{4\pi \over k} (\bar\phi_i T^a_{R_i} \phi_i) T^a$ (where $T^a$ are the
generators of the group in the fundamental representation normalized so that $\Tr(T^a T^b)
= \half \delta^{ab}$). Integrating out also $\chi$, the action takes the form (in components)
% JM Added a reference
\refs{\csntworefs,\gaiottoyin}
\eqn\actionntwo{\eqalign{S^{{\cal N}=2} = \int & {k\over {4\pi}}
\Tr(A \wedge dA + {2\over 3} A^3) + D_{\mu} \bar\phi_i D^{\mu}
\phi_i + i \bar\psi_i \gamma^{\mu} D_{\mu} \psi_i \cr &
  - {16\pi^2 \over k^2}
(\bar\phi_i T^a_{R_i} \phi_i) (\bar\phi_j T^b_{R_j} \phi_j) (\bar \phi_k
T^a_{R_k} T^b_{R_k} \phi_k) -
{4\pi \over k} (\bar\phi_i T^a_{R_i} \phi_i) (\bar\psi_j T^a_{R_j} \psi_j) \cr & -
{8\pi \over k} (\bar\psi_i T^a_{R_i} \phi_i) (\bar\phi_j T^a_{R_j} \psi_j).}}
Classically, this action includes only marginal couplings in $2+1$ dimensions, and
it has been argued \gaiottoyin\ that it cannot be renormalized beyond a possible one-loop shift of
$k$ (based on the integrality of $k$) so that it is conformally invariant also at the
quantum level.

A simple generalization of the theories above has ${\cal N}=3$ supersymmetry \refs{\kao,\cscoefficient}.
 To obtain
this we need to begin with the field content of an ${\cal N}=4$ supersymmetric theory,
namely we need to add to the vector multiplet an additional auxiliary chiral multiplet
$\varphi$ in the adjoint representation, and we need to assume that the chiral multiplets come
in pairs $\Phi_i, \tilde{\Phi}_i$ in conjugate representations of the gauge group (together
these form a hypermultiplet). The
action then includes the usual ${\cal N}=4$ superpotential $W = \tilde{\Phi}_i \varphi \Phi_i$,
and the Chern-Simons term (which breaks the ${\cal N}=4$ supersymmetry to ${\cal N}=3$)
includes an additional superpotential $W = -{k\over {8\pi}} \Tr(\varphi^2)$. Since there is
no kinetic term for $\varphi$ it can simply be integrated out, leading to a superpotential
\eqn\wnthree{W = {4\pi \over k} (\tilde{\Phi}_i T^a_{R_i} \Phi_i) (\tilde{\Phi}_j T^a_{R_j} \Phi_j).}
So, the action is the same as \actionntwo\ above (including the same terms for the conjugate chiral
multiplets),
with the addition of this marginal superpotential, whose
coefficient is determined by the ${\cal N}=3$ supersymmetry, so that it is also not
renormalized.

Another
%One
way to obtain the ${\cal N}=3$ SCFTs described above is to start
from the ${\cal N}=4$ supersymmetric Yang-Mills gauge theory with
the same matter content, and to deform it by adding a Chern-Simons
term for the vector multiplet. Using a standard normalization for
the kinetic terms of the vector multiplet with a $1/g_{YM}^2$ in
front, all components of the ${\cal N}=4$ vector multiplet then
have mass $m = g_{YM}^2 k / 4 \pi$. At low energies, compared to
this mass scale, these fields may be integrated out, leading to
the effective action described above. In the adjoint
supermultiplets we have three massive fermions with spin $+1/2$
and one with spin $-1/2$. When we integrate them out we shift the
Chern-Simons level by $k\to k-N$, but this cancels with the shift
that comes from the contribution of the gauge field. In the end
the Chern-Simons level is not changed, see \cscoefficient\ for a
more detailed discussion.

Theories with ${\cal N}$ superconformal symmetries in $d=2+1$ dimensions have an $SO({\cal N})$
R-symmetry, which in our case is $SO(3)$, or more precisely $SU(2)_R$. In the vector
multiplet, the fermions are a triplet and a singlet of $SU(2)_R$,
and the three scalar fields ($\sigma$ and two from
$\varphi$) are a triplet (as are the three auxiliary fields). In the chiral
multiplets, all fields are doublets of $SU(2)_R$. For instance,
the lowest component of $\Phi_i$, together with the complex conjugate of the lowest component
of $\tilde{\Phi}_i$ (which is in the same representation of the gauge group as $\Phi_i$) form
a doublet of this symmetry. In addition, we have a $U(N_f)$ flavor symmetry whenever
we have $N_f$ matter fields in the same representation of the gauge group.

As we reviewed, Chern-Simons theories with ${\cal N}=3 $
supersymmetry can be written rather generally. Theories with
${\cal N}=4$ supersymmetry were recently constructed in \GaiottoSD
, see also \HosomichiJD\
% JM Added this footnote with a reference to my paper with Hai Lin
 \foot{Theories with ${\cal N}=4$
  but a modified superalgebra were written in \linjm .}.

\subsec{A special case with ${\cal N}=6$ supersymmetry}

In this paper we will focus on a special case of the ${\cal N}=3$ construction above, with gauge
group $U(N)\times U(N)$ (or $SU(N)\times SU(N)$), and with two hypermultiplets in the
bifundamental representation; we will denote the bifundamental chiral superfields by $A_1, A_2$
and the anti-bifundamental chiral superfields by $B_1, B_2$. We use a notation
 similar to the one in
 the Klebanov-Witten theory \KlebanovHH\ since our field content is the same (albeit in one lower dimension),
and some of the interactions will also be similar to that theory. We will also choose the
Chern-Simons levels of the two gauge groups to be equal but opposite in sign.
Before integrating out the
vector multiplet, the superpotential in this theory takes the form
\eqn\wnsix{W = {k\over {8\pi}} \Tr(\varphi_{(2)}^2 - \varphi_{(1)}^2) + \Tr(B_i \varphi_{(1)} A_i)
+ \Tr(A_i \varphi_{(2)} B_i).}
Integrating out the auxiliary fields $\varphi_{(i)}$ (as in \KlebanovHH, when we think about
that theory as arising by flowing from a four dimensional ${\cal N}=2$ gauge theory),
we get a superpotential (which is just \wnthree\ for this special case)
\eqn\newwnsix{W = { 2 \pi \over k} \Tr(A_i B_i A_j B_j - B_i A_i B_j A_j) =
{4\pi \over k} \Tr(A_1 B_1 A_2 B_2 - A_1 B_2 A_2 B_1).}
As described above, we can obtain this theory by starting from the $2+1$ dimensional
${\cal N}=4$ gauge theory with the same field content, deforming it to an ${\cal N}=3$ theory
 by the ${\cal N}=3$ supersymmetric Chern-Simons
term, and flowing to low energies compared to the mass scale $m= g_{YM}^2 k / 4 \pi$.

Naively, the theory discussed above has only an $SU(2)$ flavor symmetry rotating the $A$'s
and $B$'s together (we will discuss additional $U(1)$ flavor symmetries below).
However, the superpotential \newwnsix\ actually has a bigger flavor symmetry, since it
can be written as
\eqn\nwnsix{W = { 2 \pi \over k} \epsilon^{ab} \epsilon^{\dot{a}\dot{b}} \Tr(A_a B_{\dot{a}}
A_b B_{\dot{b}}),}
which exhibits explicitly an $SU(2)\times SU(2)$ symmetry acting separately on the
$A$'s and on the $B$'s (as in \KlebanovHH). All other terms in our action \actionntwo\
obviously have this bigger symmetry, so we conclude that the full ${\cal N}=3$
Chern-Simons-matter theory in this special case has this enhanced global symmetry.

However, as we discussed above, these theories also have an $SU(2)_R$ symmetry under
which the bosonic fields
$A_1$ and $B_1^*$ transform as a doublet (as do $A_2$ and $B_2^*$)\foot{In the previous paragraph
$A_i$ and $B_i$ were denoting chiral superfields. In this paragraph they  denote bosonic
fields which are  the
$\theta = \bar \theta =0$ components of the chiral superfields.
  Hopefully
this will not cause confusion.}. Obviously, this
symmetry does not commute with the $SU(2)\times SU(2)$ symmetry of the previous paragraph.
The two symmetry transformations together generate an $SU(4)$ symmetry, under which the
four bosonic fields
$C_I \equiv (A_1, A_2, B_1^*, B_2^*)$ transform in the ${\bf 4}$ representation. Since $SU(2)_R$ is
an R-symmetry, we see that the supercharges cannot be singlets under this $SU(4)$. Since
general $d=3$ SCFTs have an $SO({\cal N})$ R-symmetry (with the supercharges in the
fundamental representation), we see that we must have at least ${\cal N}=6$ supersymmetry.
In the way that we wrote down the action, only the $SU(2)\times SU(2)$ subgroup of the
$SO(6)_R$ symmetry is manifest; note that the $SU(2)_R$ transformation mixes the scalar
potential coming from the superpotential $W$ with the scalar potential terms in
\actionntwo. But the full theory has an ${\cal N}=6$ superconformal symmetry, which could
be explicitly written down by performing $SU(2)\times SU(2)$ transformations on the
generators of the original ${\cal N}=3$ superconformal symmetry. In appendix A.1 we
explicitly verify that the full scalar potential is indeed invariant under $SO(6)_R$.

The full global symmetry of the theories we wrote down is
$SO(6)_R\times U(1)_b$. In the $SU(N)\times SU(N)$ case the
$U(1)_b$ symmetry is just the usual baryon number symmetry, under
which the $A_i$ have charge $(+1)$ and the $B_i$ have charge
$(-1)$. In the $U(N)\times U(N)$ theory the baryon number symmetry
is gauged by a gauge field $A_b$, but there is also an extra
gauged $U(1)_{\tilde b}$ symmetry which only couples through an
$k A_b\wedge F_{\tilde b}$ type coupling. Thus, this theory also has
a $U(1)$ global symmetry which is generated by the current $J = {
k \over 4 \pi } *F_{\tilde b}$, and the equation of motion of
$A_b$ sets this current to be exactly equal to the $U(1)_b$
current described above. (Naively there is also a global symmetry
coming from $*F_b$, but the equation of motion of $A_{\tilde b}$
implies that this acts trivially.)
 Note that in this case the flux quantization condition on $F_{\tilde b}$
 implies that the corresponding
charges are all integer multiples of $k$. So we should really say that in this case
the $U(1)$ current
is $\tilde J = J/k$.

The coupling constant of the field theories
we wrote is $1/k$, so that for large $k$ they are weakly coupled. However, in the large $N$
limit with $N/k$ fixed, as in the 't Hooft limit of standard gauge theories with adjoints
and bifundamentals \tHooftJZ,
there is an expansion in $1/N^2$ (whose leading order is given by planar diagrams), and the
effective coupling constant in the planar diagrams is the 't Hooft coupling $\lambda \equiv N/k$. Thus, the theories we discuss
are weakly coupled for $k \gg  N$, and are strongly coupled when $k \ll N$. We will see below
that in the strongly coupled regime, these theories have an alternative (dual) description that is
weakly coupled (when $N \gg 1$). In the 't Hooft limit the dual is a weakly coupled string theory.

Finally, let us note that the theory is invariant under a parity symmetry which also exchanges the
 two gauge groups, and
% OA - minor change
acts as charge conjugation on the fields,
%does a charge conjugation of the fields,
$C^I \to ( C^I)^\dagger $
(and similarly for the fermions).

\subsec{The moduli space of the $U(N)\times U(N)$ theories}

We suspect that with
this amount of supersymmetry the classical moduli space does not  receive
any quantum corrections. We begin by analyzing the moduli space in the case $N=1$. In
this case the superpotential \newwnsix\ vanishes, and so do the last two lines in
\actionntwo . (We have in this case $\sigma_{(1)} = \sigma_{(2)} =
(2\pi / k) (|A_i|^2 - |B_i|^2)$, and all the couplings involve $\sigma_{(1)}-\sigma_{(2)}$.)
Thus, this theory is simply a free theory of 4 superfields $C_I$ (two of which are
chiral superfields and two are anti-chiral superfields), with two
gauge transformations acting as $A_{(1)} \to A_{(1)} - d\Lambda_{(1)}$,
$A_{(2)}\to A_{(2)} - d\Lambda_{(2)}$, $C_I \to e^{i(\Lambda_{(1)}-\Lambda_{(2)})} C_I$,
and with a Chern-Simons term
\eqn\scs{S_{CS} = {k\over 4\pi} \int (A_{(1)} \wedge dA_{(1)} - A_{(2)} \wedge dA_{(2)}).}

We can analyze the effect of the gauge transformation on the moduli space in two
different ways. First, we can gauge-fix the gauge fields to zero. Naively we then obtain
a moduli space which is just ${\bf C}^4$. However, this gauge-fixing leaves gauge
transformations in which the $\Lambda$'s are constant everywhere unfixed. Generally, in
the presence of a boundary, such gauge transformations are not symmetries of the action,
since
\eqn\dscs{\delta S_{CS} = {k\over 2\pi} \int_{{\rm boundary}} (\Lambda_{(1)} \wedge
F_{(1)} - \Lambda_{(2)} \wedge F_{(2)}).}
However, the gauge field strengths are quantized, $\int F_{(i)} \in 2 \pi {\bf Z}$ for
an integral over any closed 2-manifold. Thus, for the action to transform by
$2\pi$ times an integer in a general configuration, we must have $\Lambda_{(i)} =
2\pi n / k$ for some integer $n$. Dividing by these additional identifications, we
obtain that the moduli space is actually ${\bf C}^4/{\bf Z}_k$, where the
${\bf Z}_k$ symmetry acts as $C_I \to e^{2 \pi i / k} C_I$. Thus, in this case our
conformal theory is simply the supersymmetric sigma model on this orbifold, which has
a manifest $SU(4)$ symmetry.

Equivalently, we can derive the same result as follows (as done in
a similar context in \dmpv).
The Abelian Chern-Simons term can be written as
$S_{CS} = {k\over 4\pi} \int A_b \wedge
F_{\tilde b}$, where $F_{\tilde b} = d A_{\tilde b}$,
$A_b = A_{(1)} - A_{(2)}$, and $A_{\tilde b} = A_{(1)} + A_{(2)}$.
The field
$A_{\tilde b}$ only appears in the action in this Chern-Simons
term, so we can dualize it into a scalar by treating $F_{\tilde b}$ rather
than $A_{\tilde b}$ as the basic variable, and adding a Lagrange
multiplier $\tau$ with \eqn\ssigma{S_{\tau} = {1\over {4\pi}} \int
\tau(x) \epsilon^{\mu \nu \lambda} \del_{\mu} F_{{\tilde b} \nu
\lambda}.} The quantization condition on $F_{\tilde b}$ (which is
the same as above) implies that $\tau$ is periodic with a period
$2\pi$. The equation of motion of $F_{{\tilde b} \mu \nu}$ implies
that $A_{b \mu} = (1/k) \del_{\mu} \tau$, and the kinetic terms of
the scalar fields in $C_I$ become $|D_{\mu} C_I|^2 = |\del_{\mu}
C_I + {i\over k} C_I \del_{\mu} \tau|^2$. Gauge invariance now
implies that the remaining gauge transformation $C_I \to e^{i
\theta(x)} C_I$ takes $\tau \to \tau - k \theta(x)$. If we now
gauge-fix this remaining transformation by taking $\tau=0$, we see
that we can still
% OA - minor change
perform
%do
gauge transformations with $\theta = 2\pi /
k$. Thus, again we obtain a sigma model on ${\bf C}^4 / {\bf
Z}_k$.

The generalization to the $U(N)\times U(N)$ case is straightforward. Whenever the
$N\times N$ matrices $C_I$ are all diagonal, it is easy to see that the scalar potential
vanishes exactly, just like in the previous case. And, it is easy to check that for
generic diagonal elements of these matrices, all off-diagonal elements are massive, so these diagonal
configurations are the full moduli space of the theory. Requiring that all the matrices
are diagonal breaks the gauge symmetry to $U(1)^N\times U(1)^N\times S_N$,
where the $S_N$ permutes the diagonal elements of all the matrices (for generic eigenvalues
only a $U(1)^N$ subgroup of this which does not act on the eigenvalues remains unbroken). Up to the
permutation symmetry we simply obtain $N$ copies of the $U(1)\times U(1)$ theory, with
the same flux quantization conditions as before for each $U(1)$ factor in the low-energy
theory (note that the $i$'th element of the fundamental representation of $U(N)$ carries
charge one under the $i$'th $U(1)$ and no charges under any other $U(1)$'s).
Thus, the moduli space in this case is simply $({\bf C}^4 / {\bf Z}_k)^N / S_N$.

Note that this is the same as the moduli space of $N$ M2-branes
probing a ${\bf C}^4 / {\bf Z}_k$ singularity in M-theory. This
theory also has ${\cal N}=6$ supersymmetry since the orbifold
preserves an $SU(4)\times U(1)$ isometry symmetry out of the full
$SO(8)$, and the ${\bf 8}_c$ representation of $SO(8)$ which the
spinors live in decomposes into $SU(4)\times U(1)$ representations
as ${\bf 6}_0 + {\bf 1}_2 + {\bf 1}_{-2}$
\refs{\nilssonpope,\susywosusy}~\foot{ This is correct for one
sign of the M2-brane charge. For the other sign of the M2-brane
charge the spinor decomposes as ${\bf 4}_1+{\bf 4}_{-1}$, so there
is no supercharge that is a singlet under the $U(1)$ and we do not
preserve any supersymmetry if $k>1$ \nilssonpope .}. For $k > 2$
the last two supercharges are projected out when we divide by
${\bf Z}_k$ in $U(1)$, so we remain precisely with ${\cal N}=6$
supersymmetry, while for $k=1,2$ the M2-brane theory has a larger
${\cal N}=8$ supersymmetry.
We conjecture that our Chern-Simons-matter theory is exactly
the same as the theory of M2-branes on the orbifold.
In particular for $k=1$ it describes M2-branes in flat space,
and for $k=2$ it describes M2-branes probing an ${\bf R}^8/{\bf Z}_2$ singularity;
% OA - minor change
in these
%in which
cases there is an enhanced ${\cal N}=8$ supersymmetry.
We have shown that the moduli spaces
% OA - minor addition
of these theories
are identical, and we
will provide further evidence for this conjecture below.

For $N=1$ the conjecture was proven above.
In particular, for $k=1$ and $k=2$ we obtain the supersymmetric sigma models on
${\bf C}^4$ and on ${\bf C}^4/{\bf Z}_2$, respectively. These sigma models, which also
arise as the low-energy theory on a single M2-brane in flat space or at an
${\bf R}^8/{\bf Z}_2$ singularity, have a larger ${\cal N}=8$ superconformal symmetry
(and a corresponding $SO(8)$ R-symmetry) which is not directly visible in the
$U(1)\times U(1)$ action that we wrote down for these theories.

Another interesting point is that if we give a vacuum expectation value to one of
the fields of the form
% OA - added factor of k^{1/2} to C_I
 $C_I = (\Lambda k)^{1/2} I_{N \times N}$,
 then the theory around
 this vacuum has an unbroken $U(N)$ gauge symmetry.
At energy scales of order $\Lambda$ we transition from the conformal regime to the moduli
space approximation. One can show, as in \dtwomtwo, that the $U(N)$ gauge fields actually become
dynamical with $g_{YM}^2 \sim \Lambda/k$, and that at energy scales below $\Lambda$
the theory reduces to the maximally (${\cal N}=8$) supersymmetric $U(N)$ gauge theory.
This description is weakly coupled for a range of energies when $g_{YM}^2 N \ll
\Lambda$, or $k \gg N$.
This is also clear from the picture of
the branes probing the $\bf{C}^4/\bf{Z}_k$ singularity. If the M2-branes are sitting away
from the origin, then for large $k$ the ${\bf Z}_k$ identification looks like an identification on
a small circle transverse to the M2-branes, and
there is an energy range where the theory reduces to the
${\cal N}=8$ super Yang-Mills theory of D2-branes. In terms of the brane probes the
configuration is similar to  the
 ``deconstruction'' configuration in \nimadec . An important difference is that here
  we do not increase the number of dimensions
 where the theory is defined.

\subsec{Chiral operators and Wilson lines}

From the point of view of ${\cal N}=2$ supersymmetry, we can
obtain chiral primary operators by taking any gauge-invariant
products of $A$'s and $B$'s, modulo the $F$-term equations. In
particular, we can take operators of the form $\Tr((A_a B_{\dot a})^l)$
for $l=1,2,\cdots$, in which the $SU(2)\times SU(2)$ indices are multiplied
symmetrically (since anti-symmetric combinations vanish in the
chiral ring due to the $F$-term equations). In the ${\cal N}=6$
language, the $A$'s are part of a ${\bf 4}$ of $SU(4)_R$, and the
$B$'s are part of a ${\bf \overline 4}$, so these operators are a
subset of operators in the $l$'th symmetric product of ${\bf 4}$'s
times the $l$'th symmetric product of ${\bf \overline 4}$'s, with no
contractions between ${\bf 4}$'s and ${\bf \overline 4}$'s. More precisely,
the Dynkin labels of the resulting $SU(4)$ representation are $(l,0,l)$.
This full class of operators, which we can write schematically as
$\Tr((C_I C^{\dagger}_J)^l)$, must be chiral in the full ${\cal
N}=6$ theory. The scaling dimension of these operators is
$\Delta=l$. None of these operators carry any charge under
$U(1)_b$.

How can we obtain any
% OA - minor change
operators
%states
carrying $U(1)_b$ charge ? Recall
that in the $U(N)~\times~U(N)$ theories
this charge is generated by $J=(k/4\pi) *F_{\tilde b}$, so
% OA - minor change
the corresponding states
%these states
must involve a non-zero magnetic flux in the diagonal
$U(1)$ gauge group. This is a flux on the $S^2$ surrounding the
point where the operator is inserted.  These operators are called
't Hooft operators or monopole operators and were discussed in detail in
\refs{\kapustinthooftone,\kapustinthoofttwo}.
Let us begin by discussing
this in the $U(1)\times U(1)$ case. A configuration with $n$ units
of flux in each of the $U(1)$'s has $nk$ units of charge under
$J_b$, and the equation of motion of $A_b$ tells us that this
configuration must also carry $(-n k)$ units of the ``original''
baryon number charge (given by the number of $A$'s minus the
number of $B$'s). In the $U(1)\times U(1)$ theory, we can thus
obtain additional gauge-invariant operators by taking a product of
$n k$ $B_i$'s, and adding $n$ units of flux in each $U(1)$. It is
also useful to understand how these states arise when we consider
the theory on $S^2 \times {\bf R}$. In this case we can construct
a well defined state as follows. The $B$ fields are now massive
(due to their conformal coupling to the curvature) and their
lowest modes are created by harmonic oscillator creation
operators. We can consider a state that has $q$ oscillators
excited, which would carry $-q$ units of the original baryon
charge. These states carry charges $(-q,q)$ under the  $U(1)
\times U(1)$ gauge fields. We can now also put in $n$ units of
flux for each $U(1)$ gauge field on the $S^2$. Due to the Chern
Simons action this flux configuration has baryon number charge $ n
k$. Thus we see that if we set $q= nk$ we get an invariant
configuration. More explicitly, the equations of motion for the
gauge fields have the form \eqn\eqmotgf{
 { k \over 2 \pi } F_{(1)} - *  { i \over 2} (   \bar B d B -  B d { \bar B} )  =0 ~,~~~~~~~- { k \over 2 \pi}
 F_{(2)} + *  { i \over 2} (   \bar B d B -  B d { \bar B} )  =0~.
 }
 Integrating this on $S^2$ in a configuration where $\int_{S^2} F_{(1)} = \int_{S^2} F_{(2)} = 2\pi n$ and
 where the total charge carried by the field $B$ is $-q$ we see that both equations in \eqmotgf\ reduce
 to $k n =q$.
 A third way to think about this problem
is the following. The operator $B^{nk}$ is charged under the
$U(1)_b$ gauge symmetry. So we can attach a Wilson line that ends
on it and which couples to the $U(1)_b$ gauge field. This is a
Wilson line with charges $(nk,-nk)$ under $U(1) \times U(1)$. Thus,
we can think of an operator of the form $  ( e^{ i k
\int_{\infty}^0 A } )^n  B^{n k}(0)$. This operator appears to be
non-local. However, due to the Chern-Simons terms, such a Wilson
line is not observable since it is equivalent to a large gauge
transformation with gauge parameters $ ( \epsilon_{(1)} ,
\epsilon_{(2)} ) = (n \varphi,n \varphi) $ where $\varphi $ is the
angle around the Wilson line, see \mooreseiberg . Then we conclude
that we have a local operator.
 An equivalent description is in terms of $n$ 't Hooft
disorder operators \tHooftHY\ inserted at the position where we insert the
field $B^{nk}$. These are operators which insert a flux of magnetic
field on a sphere surrounding the point where the operator is
inserted. In the language we used in our discussion around \ssigma, we can
write these operators as $e^{in\tau(x)}$. In this language the full operator is
$e^{ i n \tau} B^{n k }$.

 Finally, we conclude that
  the spectrum of chiral operators in this theory includes, in
addition to the operators described in the first paragraph, operators formed from
$n k$ $A_i$'s (or $n k$ $B_i$'s), plus fluxes or unobservable
Wilson lines ending on them.  Generalizing this using the
$SU(4)_R$ symmetry we deduce the existence of gauge-invariant
operators of the form $C^{nk}$ (or $(C^{\dagger})^{nk}$) which are
in the $(n k)$'th symmetric product of ${\bf 4}$'s (${\bf
\overline 4}$'s) and carry $n k$ units ($-n k$ units) of $U(1)_b$
charge.

%\ifig\wilson{  A local operator constructed from a symmetric product of $k$ complex
%fields $C$ at the origin
%plus a Wilson line that goes from infinity to the origin. The operator is local because
%the Wilson line in the   $ (  {\rm Sym }({\bf N}^{k}), {\rm Sym }(\overline {\bf N}^{k}) )  $
%representation is not observable.  } {\epsfxsize1.5in\epsfbox{wilson.eps}}

% JM
In the $U(N)\times U(N)$ case
we can perform a similar construction
\refs{\tHooftHY\GNO-\kapustinthooftthree}. The end result is that operators of the form
$C^{n k}$ are allowed. Naively such operators would not be gauge-invariant.
They would transform in the symmetric product of $nk $ fundamentals
of the first $U(N)$ and $nk$ anti-fundamentals of the second $U(N)$.
Let us first consider the operator $C^k$. The non-Abelian theory also contains
't Hooft operators, or monopole operators \refs{\GNO\kapustinthooftthree-\kapustin}.
We can consider such operators with one unit of flux on each of the two $U(N)$ gauge groups.
In the presence of the Chern-Simons term such an operator transforms in the $
 (  {\rm Sym }({\bf N}^{k}), {\rm Sym }(\overline {\bf N}^{k}) )  $ representation of
 $U(N) \times U(N)$. Thus it can be combined into a singlet with $C^k$. For the case $C^{nk}$
 we can consider 't Hooft or monopole operators with $n$ units of flux on each $U(N)$ factor\foot{
 The previous version of this paper stated that one could also think of these operators in terms
 of Wilson lines that end at the insertion point. Though that is correct in the Abelian case,
 and for Chern-Simons theories
 with no charged matter \ItzhakiRC , it appears to be incorrect in our non-Abelian case when there is charged matter.
  We thank M. Schnabl for
 pointing this out to us.}.
Notice that for the particular case of $k=1$ we can construct in this way chiral primary operators
with a single complex field $C$. These operators have dimension $\Delta = 1/2$ which
saturates the unitarity bound, and should thus be free fields. In the dual description,
these operators and their complex conjugates
describe the center of mass motion of the M2-branes. For $k=1,2$
one can
construct dimension two currents using 't Hooft operators. By virtue of their dimensions,
these are conserved and they enhance the global symmetry to $SO(8)$
\foot{ We thank E. Witten for this argument.}.

Alternatively, we can note that if we quantize the theory on $S^2 \times {\bf R}$ we can diagonalize
the oscillator $B $ and go to a highest weight state where only the oscillator corresponding to the
first eigenvalue is excited $k$ times. We can then do an analysis similar to what we did on
 the moduli space and in the $U(1)\times U(1)$ case, and
conclude that this charge can be canceled by a flux which is only in the first
$U(1)$ factor.
% \times U(1)$ factor.
We can view the resulting state as a lowest weight state of the representation which is the symmetric product of
$k$ fundamentals for the first $U(N)$ factor and of $k$ anti-fundamentals for the second.

Finally, let us give a more complete description of the chiral
primary operators. For this purpose let us concentrate on the
chiral primary operators under an ${\cal N}=2$ subalgebra of the
full supersymmetry algebra. The chiral primary operators are made
with the fields $A_i$ and $B_j$ subject to the relations that
follow from the derivatives of the superpotential \nwnsix . In
addition we can construct operators of the form $ A^k$ using the 't Hooft (or monopole) operator
described above. Of course
we can also combine them into operators of the rough form $A^{nk}
(AB)^l$, $B^{nk} (AB)^l$. In such operators there are many ways to contract the
indices and only some combinations will   give protected operators. We can find
the protected chiral  operators by noticing that
these operators are given by considering a certain quantum
mechanics on the moduli space, as in \refs{\indexref\martelli-\hananyexp}. The
moduli space is ${ \bf C}^4/{\bf Z_k}$. So we have $N$ bosons on
this space. More precisely, we imagine that the single particle
Hilbert space is that of four harmonic oscillators subject to the
constraint that all states have zero ${\bf Z_k}$ charge. We then
consider $N$ particles on this moduli space. This
describes the large $N$ limit of the ${\cal N}=2$ chiral ring. In
addition we might have to add some operators of baryonic type, see
\hananybaryon . We will ignore these extra operators for the time
being. The final conclusion is that the ``single-trace'' operators are
given by the one-particle Hilbert space we described above.
Once we take into account the full $SO(6)$ symmetry the full
space is the  same as the space of  $SO(8)$ spherical
harmonics that are invariant under ${\bf Z}_k$.  Of course, we also
have the full Fock space constructed from products of these.\foot{
This discussion of the spectrum of general chiral primary
operators that are charged under the $U(1)$, which followed \hananyexp ,
is based on the moduli space. Thus, matching the spectrum of
chiral primary operators that we found (at large $N$) to the dual
field theory of M2-branes is not really an independent test of the duality.
It would be nice to perform a
direct operator computation of the chiral ring, in order to display more
explicitly the operators in the chiral ring.
}

Note that the spectrum of chiral operators becomes
$SO(8)$-invariant for $k=1,2$. The $SO(8)_R$  symmetry mixes the standard
operators described at the beginning of this section with the
non-standard operators which include 't Hooft disorder operators.
 This suggests that this symmetry (and, thus,
the ${\cal N}=8$ superconformal symmetry) does not act locally on
the fields in our Lagrangian. This is related to the fact that the
transformation between our description of these theories and a
description in which the ${\cal N}=8$ supersymmetry is manifest
involves a mirror symmetry transformation in three dimensions (as
we will discuss in the next section), which is generally
non-local.

In the $SU(N)\times SU(N)$ theory we have additional baryon-like (or di-baryon)
  chiral operators of
the form $\det(C)$ \GubserFP\ (with various flavor indices for the $C$'s),
%with the gauge indices of both groups contracted by epsilon symbols,
carrying $N$ units of $U(1)_b$ charge. In the $U(N)\times U(N)$
theory these operators might  not be gauge-invariant, but if we
take a product of $n$ of these operators, with $n N$ a
multiple of $k$, we can form a gauge-invariant operator by adding
flux as described above. In fact, in this case there is no real distinction
between these operators and the gauge-invariant operators
described in the previous paragraphs.

Additional interesting operators are Wilson lines. Since we have bifundamentals
some Wilson lines can be easily screened by creating excitations of the $C$ fields.
If we consider Wilson lines which contain $n$ copies of the fundamental representation of
the first $U(N)$ factor then generically they cannot be screened. However, if we take $n=k$,
then
the Wilson lines can be screened or
 become unobservable because they are equivalent to doing a large
gauge transformation. More specifically, we can see that if we have a compact space, such
as $S^2 \times {\bf R}$ and we align $n$ Wilson lines in the fundamental representation of
one of the gauge groups along
the time direction, then the observable is zero for $n< k$ due to the Gauss law.
On the other hand, it can have a non-zero value for
$n=k$. In this case we can add a flux on the $S^2$ and the configuration can become gauge-invariant.

Similarly, if we consider  $N$ Wilson lines, they can combine into a singlet under $SU(N)$.
In the $U(N)$ theory
they would still carry charge under the overall $U(1)$. This charge can be screened by flux
if $N$ is a multiple of $k$, or, if we have $n N$ Wilson lines, when $n N$ is an integer multiple of $k$.

\subsec{The $SU(N)\times SU(N)$ theories}

The moduli space for the $SU(N)\times SU(N)$ theories is slightly more complicated
than the $U(N)\times U(N)$ case described above. The moduli space is still described
by diagonalizing all four matrices $C_I$, but the identifications are more complicated.
In the $U(N)\times U(N)$ case we had independent $U(1)\times U(1)$ gauge transformations
acting on the $j$'the eigenvalue of $C_I$, $C_I^j$ ($j=1,\cdots,N$), as $C_I^j \to
 e^{i(\Lambda_{(1)}^j - \Lambda_{(2)}^j)} C_I^j$, and the arguments around \dscs\
suggested that we should identify configurations where the $\Lambda$'s are independently
integer multiples of $2\pi/k$ (in addition to the permutation of the eigenvalues). In the
$SU(N)\times SU(N)$ case the $\Lambda$'s are constrained by $\sum_j \Lambda_{(1)}^j =
\sum_j \Lambda_{(2)}^j = 0$. The effective Chern-Simons term on the moduli space
looks like $N$ copies of \scs, but with the constraint $\sum_j A_{(1)}^j =
\sum_j A_{(2)}^j = 0$. If we consider the variation of the Chern-Simons term as in
\dscs, we find that the gauge transformations that we need to identify by are the
ones for which $k \sum_j \Lambda_{(1)}^j f_j \in 2 \pi {\bf Z}$ for any integers $f_j$
obeying $\sum_j f_j = 0$, since these are the allowed fluxes in the $SU(N)$ theory (and
similarly for the $\Lambda_{(2)}$'s). Defining
$\omega \equiv \exp(2 \pi i / N k)$, a
basis for these identifications is provided by the identifications
$g_l : C_I^j \to \omega^{1-N \delta_{jl}} C_I^j$ for $l=1,\cdots,N-1$, which are all
consistent gauge transformations of the type described above\foot{An alternative basis
is provided by the elements $g_1$ and $h_l = g_l g_1^{-1}$ ($l=2,\cdots,N-1$), where
$h_l$ acts as $C_I^1 \to \exp(2\pi i / k) C_I^1$, $C_I^l \to \exp(-2\pi i / k) C_I^l$.}.
The full moduli space
is the space $\bf{C}^{4N}$ divided by the $(N-1)$ $g_l$ identifications and by the permutation
group $S_N$ (whose elements do not commute with $g_l$). It is not clear if there is
any simple description of the full group of identifications in this case.\foot{One can
also consider the $(SU(N)\times SU(N))/{\bf Z}_N$ theory, which has a somewhat different
moduli space because of the different fluxes that are allowed.}

Naively one may think that there is a simple relation between the $U(N)\times U(N)$
and the $SU(N)\times SU(N)$ theories; the former should just arise by gauging the
global $U(1)_b$ symmetry of the latter theory, adding another free $U(1)$ gauge field, and adding appropriate supersymmetric Chern-Simons couplings for the two $U(1)$'s. This would suggest
that the moduli space of the $U(N)\times U(N)$ theory should just be an orbifold of the
$SU(N)\times SU(N)$ case, by an argument similar to the one we gave in our analysis of the
$U(1)\times U(1)$ case. However, this argument is not correct, since $U(N)$ is not simply
the product of $U(1)$ and $SU(N)$, but rather it is $(U(1)\times SU(N))/{\bf Z}_N$, such
that the flux quantization conditions are not the same for $U(N)\times U(N)$ as for
$U(1)\times SU(N)\times U(1)\times SU(N)$. This is why the moduli space we found for
the $U(N)\times U(N)$ case is not simply an orbifold of the $SU(N)\times SU(N)$ moduli space\foot{We could also take different levels for the $U(1)$ and $SU(N)$ factors in $U(N)$,
but we will not discuss this possibility here.}.

\subsec{The $N=2$ case and comparison with the Bagger-Lambert theory}

In the special case of $SU(2)\times SU(2)$, the bifundamental
hypermultiplets are in the real $\bf(2,2)$ representation, so
$A_1$, $A_2$, $B_1$ and $B_2$ are all in the same representation,
and without the superpotential the action \actionntwo\ has an
$SU(4)$ flavor symmetry. We prove in appendix A.2 that in this
case the superpotential \nwnsix\ also has this $SU(4)$ flavor
symmetry. (Note that the $3+1$-dimensional Klebanov-Witten theory
\KlebanovHH\ also has an $SU(4)$ global symmetry in the $SU(2)
\times SU(2)$ case \hananybaryon ).\foot{In the three dimensional
${\cal N}=4$ super-Yang-Mills theory with this matter content the superpotential
breaks this symmetry to an $SO(4)$ global symmetry. However, it
turns out that after flowing to the ${\cal N}=3$ Chern-Simons
theory, the superpotential \nwnsix\ actually has the full $SU(4)$
global symmetry.} This $SU(4)$ symmetry is a symmetry of the
superpotential (written in ${\cal N}=2$ notation) and it acts on
the chiral superfields. This should not be confused with the
$SU(4)_R$ symmetry that we had for general $N$. In addition, we
still have
 the $SU(2)_R$ symmetry, which exchanges the $A$'s with the
complex conjugates of the $B$'s, and  does not commute with the $SU(4)$ global
symmetry. These two symmetries combine to give
  an $SO(8)$ global R-symmetry.
This $SO(8)$ symmetry  rotates  all $8$ real scalars in the
$\bf(2,2)$ representation. As above, this implies that in this case our theories
actually have ${\cal N}=8$ superconformal symmetry (for any value of $k$), which can
be explicitly realized by
flavor transformations on the usual supercharges.
In fact, one can show
that our $SU(2)\times SU(2)$ (or $(SU(2)\times SU(2))/{\bf Z}_2$) theory is precisely the same as that of \baggerlambert\
(which was written as a
Chern-Simons-matter theory in \markvr).
%The identification by $g_1$ described above
%is then simply the $\bf{Z}_{2k}$ orbifold discussed in \refs{\dmpv,\lamberttong},
% and the full identification on the $\bf{C}^8$ moduli space
%is by the dihedral group $D_{2k}$.

For $k=1$ this theory was claimed to describe
two M2-branes on ${\bf R}^8/{\bf Z}_2$.
However, this cannot be precisely correct, since the moduli space does not include
the ${\bf Z}_2$ transformation acting on each M2-brane separately.
This claim was motivated by the Type IIA picture of two D2-branes at an
orientifold 2-plane, which lifts to the M-theory ${\bf Z}_2$ orbifold.
It was assumed that the theory on the D2-branes is the
${\cal N}=8$ $SO(4)$ super Yang-Mills theory, which would then flow to
 the $SU(2)\times SU(2)$ $k=1$  Chern-Simons theory in the IR.
However, the D2-brane gauge group is actually $O(4)$, not $SO(4)$.
The theory of two M2-branes on ${\bf R}^8/{\bf Z}_2$ should therefore
correspond to the IR limit of the ${\cal N}=8$ $O(4)$ super Yang-Mills theory.
The extra ${\bf Z_2}$ in the gauge group provides precisely the extra
${\bf Z_2}$ projection in the moduli space.
For $k=2$ the theory was claimed \refs{\lamberttong,\dmpv} to describe two M2-branes on ${\bf R}^8/{\bf Z}_2$
with discrete torsion of the 3-form field \SethiZK .
This was understood as the low energy limit of the super Yang-Mills theory
of two D2-branes at an orientifold 2-plane
with discrete RR torsion (the so-called $\widetilde{O2}^-$), which has the
gauge group $SO(5)$.

On the other hand, our analysis suggests (and more evidence will follow)
that the theory of any number $N$ of M2-branes on the orbifold
${\bf R}^8/{\bf Z}_k$ is $U(N)\times U(N)$ at level $k$.
In particular the theory of two M2-branes on ${\bf R}^8/{\bf Z}_2$ is
$U(2)\times U(2)$ at level $k=2$.
The question of discrete torsion is interesting, and we leave it for a future
investigation.

%%%%%%%%%%%%%%%%%%%%%%%%%%%%%%%%%%%%%%%%%%

\newsec{Brane constructions}
%\newsec{Brane constructions of the ${\cal N} =3$ Yang-Mills-Chern-Simons theory and its IR limit}

Three dimensional gauge theories with a Chern-Simons term can be
realized in brane constructions in Type IIB string theory
\refs{\KitaoMF,\bergman}. Generically these theories can have at most
${\cal N}=3$ supersymmetry. In this section we will generalize
these constructions to theories with a $U(N)\times U(N)$ gauge
group, with Chern-Simons terms at levels $k$ and $-k$,
respectively, and with matter in the bi-fundamental
representation. These theories flow in the IR  to precisely the
${\cal N}=6$ superconformal Chern-Simons theories considered in
section 2. On the other hand, lifting the configuration (via
T-duality) to M-theory will allow us to relate these theories (at
low energies) to
M2-branes probing a ${\bf C}^4/{\bf Z}_k$ singularity, and thereby
% OA minor addition
to
justify
the duality conjecture we made in the previous section.
%our
%conjecture about the supergravity dual in the large $N$ limit.

\subsec{Type IIB brane configurations with ${\cal N}=3$ supersymmetry}

%%% TO PUT FIGURES INSERT:
\ifig\HWconfiguration{ The brane configuration for the ${\cal N}=4$
supersymmetric theory in three dimensions with a $U(N) \times U(N)$
gauge group and two bifundamental hypermultiplets. We have
NS5-branes along 012345 and $N$ D3-branes along 0126.
  } {\epsfxsize1.2in\epsfbox{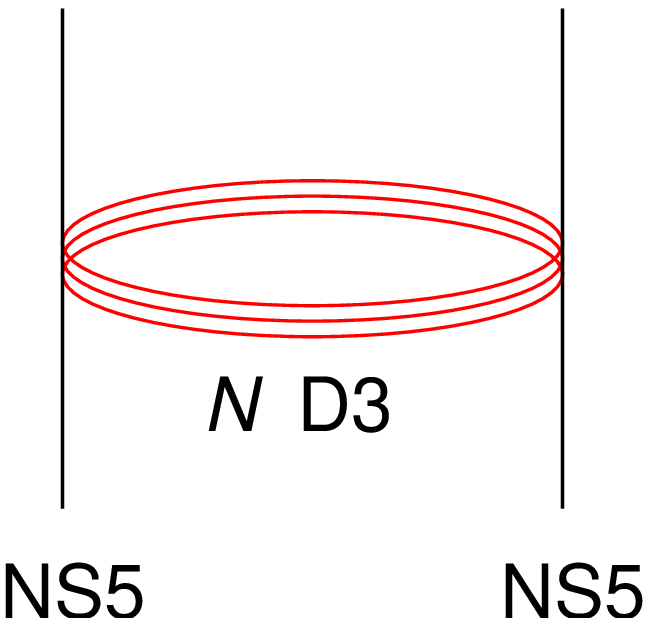}}

Our construction of the Type IIB brane configurations will follow
closely the approach of \bergman. We start with the brane
configuration for a three-dimensional ${\cal N}=4$ supersymmetric
gauge theory consisting of two parallel NS5-branes along the
directions 012345 and separated along the compact direction 6, and
$N$ D3-branes along the directions 0126 \HananyIE , see
\HWconfiguration . The directions 012 are common to all the
branes, and are identified with the coordinates of our
three-dimensional field theories. The D3-branes wind around the
compact direction 6, but since they can break on the NS5-branes we
get two $U(N)$ vector multiplets, one for each interval. In
addition, the open strings between D3-branes across the NS5-branes
give rise to two complex bifundamental hypermultiplets.
In ${\cal N}=2$ notation, the bifundamental hypermultiplets give
rise to chiral superfields $A_i$ ($i=1,2$) in the $\bf(N, \bar N )
$ representation, and $B_j$ ($j=1,2$) in the $\bf(\bar N , N)$
representation\foot{This theory can also be viewed as the
dimensional reduction of a four dimensional ${\cal N}=2$ gauge
theory which arises by a similar construction with D4-branes in
Type IIA string theory, or alternatively by considering D3-branes
at an $\bf{R}^4/\bf{Z}_2$ singularity (which is related by
T-duality to the D4-brane construction).}. At this point we have
an ${\cal N}=4$ $U(N)\times U(N)$ gauge theory, with a dynamical
gauge field, three scalars and four fermions in the adjoint
representation of each gauge group, and four complex bosons in the
$\bf(N, \bar N)$, their complex conjugates in the  $\bf(\bar N,
N)$, and their fermionic partners.

\ifig\Dfive{ We add D5 branes along 012349 to the configuration in
\HWconfiguration\ so that we get an ${\cal N}=2$ theory which is
the ${\cal N}=4$ theory of \HWconfiguration\ plus $k$ massless
chiral multiplets in the fundamental and $k$ in the
anti-fundamental of each gauge group.}
{\epsfxsize1.3in\epsfbox{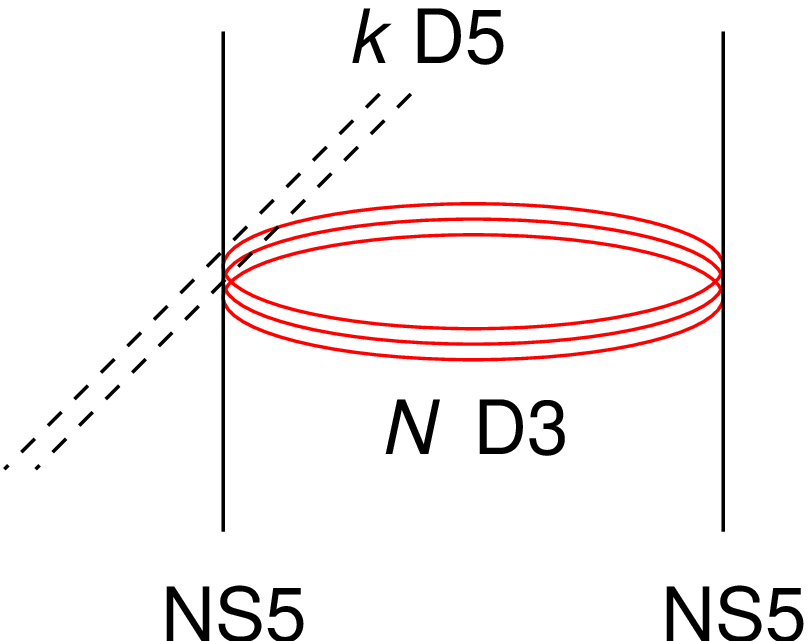}}

Next, we add $k$ D5-branes along 012349, such that they intersect
the D3-branes along 012, as well as one of the NS5-branes along
01234, see \Dfive .    This breaks the supersymmetry to ${\cal
N}=2$, and adds $k$ massless chiral multiplets in the fundamental
and $k$ massless
 chiral multiplets in the anti-fundamental representation of each
of the $U(N)$ factors.  (Note: this is different from the
D5-brane orientation considered in \HananyIE, which preserved
${\cal N}=4$ supersymmetry.) Each chiral multiplet consists of a
two-component Majorana fermion and a complex scalar. We will now
recall how to obtain the Chern-Simons term by a mass deformation
of this configuration \bergman. There are
% OA - minor change since counting is unclear
several
%four
possible mass
deformations. Separating the D5-branes from the D3-branes in the
directions 78 corresponds to a standard complex mass parameter in
the superpotential, which is inherited from the four-dimensional
${\cal N}=1$ theory. The separation in the 5 direction corresponds
to a real mass term of equal magnitude but {\it opposite sign} for
the fundamental chiral and anti-fundamental chiral multiplets.

%%% TO PUT FIGURES INSERT:
\ifig\web{The web deformation of the intersecting NS5-D5
configuration,
% OA - added clarification
as seen in the $59$ plane.}
{\epsfxsize2.5in\epsfbox{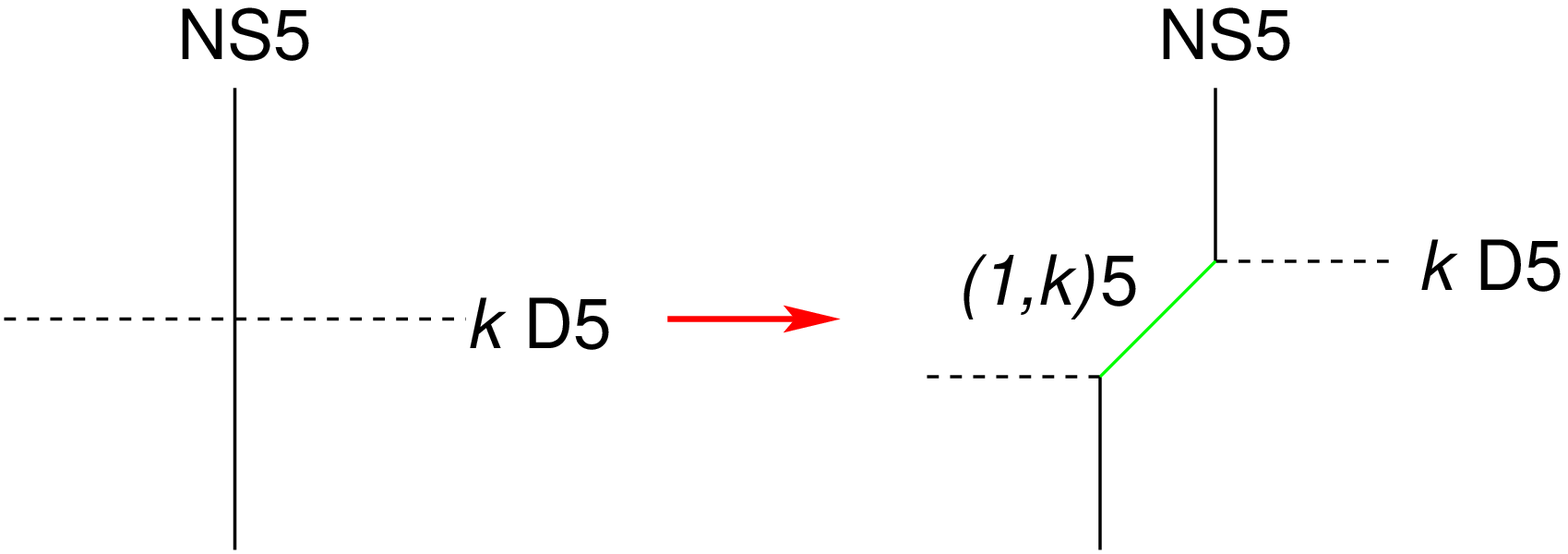}}

The final deformation,
corresponding to a real mass term of {\it equal sign} for the
fundamental  and anti-fundamental chiral multiplets,
is a {\it web deformation}, in
which the $k$ D5-branes and NS5-branes break along the directions
% OA - minor correction
01234
%1234
and merge into an intermediate $(1,\pm k)$5-brane, see \web\ \bergman . The
sign of the mass term depends on the relative sign of the charges
of the intermediate 5-brane. Supersymmetry fixes the angle of the
$(1,k)$5-brane (relative to the NS5-brane) in the 59 plane to be
\AharonyJU\ (see also appendix B)
\eqn\relang{
 \theta = \arg(\tau) - \arg(k + \tau) \;
, \; \tau = {i\over g_s} + \chi \,. } In particular for $\chi=0$
and $g_s=1$ the angle is $\tan \theta=  k$. This is the
deformation which interests us.
% OA - minor addition and subtraction of "then" below
In this case,
integrating out the fermions in
the chiral and anti-chiral multiplets
%then
produces a Chern-Simons
term via the parity anomaly \refs{\niemimics\AlvarezGaumeIG,
\RedlichDV}. The coefficient of the CS term gets a contribution of
$+1/2$ from each Majorana fermion with a positive mass term, and
$-1/2$ from each fermion with a negative mass term. We therefore
get a total coefficient $k$ for one of the $U(N)$ factors, and
$(-k)$ for the other.
% OB - Let me clarify this point a bit:
The two CS coefficients have opposite signs because the relative positions
of the NS5-brane and $(1,k)$5-brane on the second interval are exchanged
relative to the first.
Note that if we perform an $SL(2,{\bf Z})$ transformation taking $\chi \rightarrow \chi - k$,
then this changes the $(1,k)$5-brane into an NS5-brane, and the NS5-brane
into a $(1,-k)$5-brane. We would then have on the second interval the
same relative positions as on the first interval, except that $k \rightarrow -k$.

We end up with an NS5-brane along 012345 and a $(1,k)$5-brane
along $01234[5,9]_\theta$, where $[5,9]_\theta$ corresponds to the
direction $x_5\cos\theta + x_9\sin\theta$. This gives at low energies a
$U(N)_k\times U(N)_{-k}$ Yang-Mills-Chern-Simons theory with
${\cal N}=2$ supersymmetry, four massless bi-fundamental matter
multiplets and their complex conjugates, and two massless adjoint
matter multiplets corresponding to the motion of the D3-branes in the
directions 34, which are common to the two 5-branes. The vector
multiplet has a mass given by $g_{YM}^2k/(4\pi)$. The mass term
for the two adjoint fermions in the vector multiplet of each
$U(N)$ has the same sign, and the sign is opposite for the two
$U(N)$'s. Integrating out these fermions at one loop therefore
shifts the level of the Chern-Simons term by $\pm N$. However, this shift is
canceled by an opposite shift due to the massive gauge field
\cscoefficient .

\ifig\CSconfiguration{
 The final brane configuration that gives
rise to the ${\cal N}=3$ theory.   All branes are stretched along
the directions 012. The D3-branes are also stretched along a
compact direction 6, and the NS5-branes are stretched also along
the directions 345. The $(1,k)$ fivebrane  is also stretched along
directions mixing the directions 345 and 789, namely $[3,7]_\theta
[4, 8]_\theta [5,9]_\theta$. }
{\epsfxsize1.5in\epsfbox{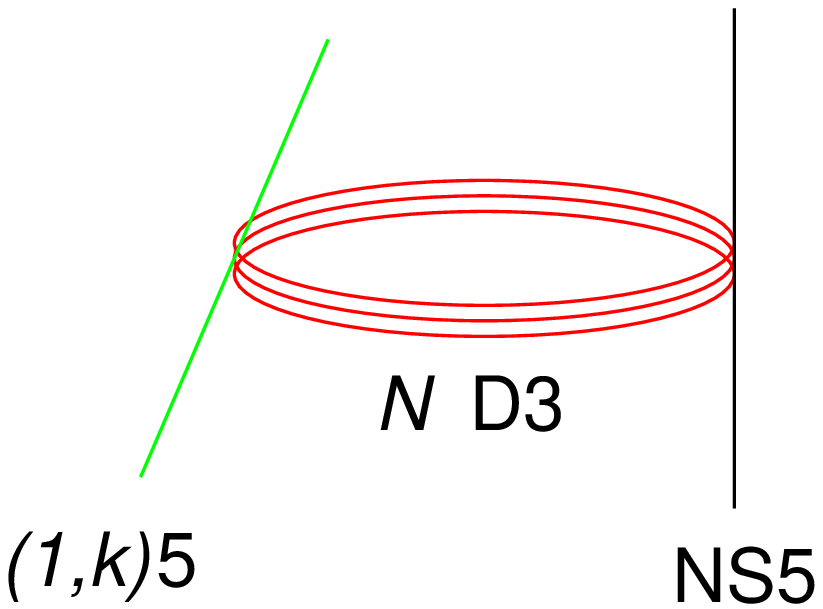}}

To get the ${\cal N}=3$ theory we first generalize the
construction by rotating the $(1,k)$5-brane relative to the
NS5-brane in the 37 and 48 planes. To preserve ${\cal N}=2$
supersymmetry the two angles must be equal. The $(1,k)$5-brane is
then along $012[3,7]_\psi[4,8]_\psi[5,9]_\theta$. This gives a
superpotential mass proportional to $\tan\psi$ to the two adjoint
matter multiplets. For the two fermions in these multiplets the
sign of the mass term is opposite, so they do not change the
level. For the special case of $\psi=\theta$ the masses of all the
adjoint fields are equal, and there is an enhanced ${\cal N}=3$
supersymmetry.
In conclusion, the brane configuration under consideration
consists of a $(1,0)$ (or NS) fivebrane along 012345, and a
$(1,k)$ fivebrane along $012[3,7]_\theta [4, 8]_\theta
[5,9]_\theta$. The angle is given by \relang .
 The two branes are separated along the direction 6,
and we have also $N$ D3-branes stretched along this compact
direction intersecting the two fivebranes, see \CSconfiguration . The
whole configuration has ${\cal N}=3$ supersymmetry. The $SO(3)_R$
symmetry corresponds to rotations by the same $SO(3)$ element in
the 345 and the 789 subspaces.
%Similar configurations were previously considered in REFERENCES.

In the brane construction it is clear that if we separate $m$
D3-branes from the fivebranes in the transverse directions, so
that they do not intersect either of the fivebranes, then at low
energies we obtain the ${\cal N}=8$ supersymmetric $2+1$
dimensional $U(m)$ Yang-Mills theory. This brane motion
corresponds to giving vacuum expectation values to the
bi-fundamentals.

\subsec{The lift to M-theory}

For simplicity, we will take the axion
% OA - minor addition
$\chi$
to vanish.
We begin by T-dualizing along the direction 6, turning it into the
direction $\tilde 6$ in type IIA string theory.
 This transforms the D3-branes into D2-branes. The NS5-brane becomes a Kaluza-Klein (KK) monopole
 along $012345$, associated with the circle $\tilde 6$ (namely, this is the circle which
 shrinks at the core of the KK monopole).
 Similarly, the $(1,k)$ fivebrane becomes an object
 that sits along $012[3,7]_\theta [4, 8]_\theta [5,9]_\theta$, which consists of a KK monopole associated with the $\tilde 6$ circle together with $k$ D6-branes.
 Of course, these D6-branes are not actual branes, but become flux
 on the KK monopole \SenZB .
 It is again clear that separating the D2-branes from the other
 objects gives the theory on D2-branes in flat space.

We can now lift the configuration to M-theory, where we get a new
direction, 10. The D2-branes become M2-branes, the KK monopole
remains a KK monopole associated with the $\tilde 6$ circle, and
the D6-brane becomes a KK monopole associated with the circle in
the 10 direction. The $(1,k)$5-brane is therefore lifted to a
single KK monopole
%Thus, the configuration that we got from the $(1,k)$ fivebrane
%(that is at an angle $\theta$ in the 345-789 directions) becomes a single KK monopole
that is associated with a circle given by a linear combination of
$\tilde 6$ and $10$. In other words the two 5-branes in the Type
IIB configuration have lifted to a pure geometry in M-theory,
which is probed by the M2-branes.
%The conclusion is that both of the original fivebranes became
%a geometrical configuration in M-theory, which is now being probed by the M2-branes.
The eleven-dimensional geometry is ${\bf{R}}^{1,2} \times X_8$,
where $X_8$ is the space we get by the superposition of the two
Kaluza-Klein monopoles. The M2-branes are extended along
$\bf{R}^{1,2}$ and probe $X_8$. This eight-dimensional space
preserves $3/16$ of the supersymmetry \kkmonopole . Adding the
M2-branes does not break any additional supersymmetry, as long as
we add them with the right orientation. Such eight dimensional
spaces were studied in detail in \kkmonopole, and we summarize
their geometries in appendix B. These spaces are a generalization
of the Gibbons-Hawking metric \gibbonshawking , and they obey a linear
superposition principle. This allows us to find the explicit
geometry for the case at hand. We simply need to add the solutions
for each of the Kaluza-Klein monopoles (see appendix B for more
details). Since the Kaluza-Klein monopoles have charge one, the
M-theory geometry is completely smooth near each of the monopole
cores. In fact, near each core it looks like $\bf{R}^4$. However,
when we look at the intersection region of the two Kaluza-Klein
monopoles, we find that the geometry is locally
$\bf{R}^8/\bf{Z}_k$ or $\bf{C}^4/\bf{Z}_k$. The $\bf{Z}_k$ acts as
$z_I \to e^{ i 2 \pi/k} z_I$ on the $\bf{C}^4$ coordinates. This
is explicitly shown in appendix B. This is the  only
singularity of the metric.

\subsec{The IR limit}

On the field theory side it is clear that at low energies the
${\cal N}=3$ Yang-Mills-Chern-Simons gauge theory that we
constructed flows to the ${\cal N}=6$ superconformal Chern-Simons
theory with bi-fundamental fields that we discussed in the
previous section. This can be seen explicitly by integrating out
all the massive fields.
%This infrared limit is simplest to study in the M-theory picture, where it
In the M-theory picture, on the other hand, the IR limit becomes
the near-horizon limit of the M2-branes probing the
$\bf{C}^4/\bf{Z}_k$ singularity. This singularity preserves 12
supercharges (although the original 8-dimensional space $X_8$
preserved only 6). The M2-branes do not break any more
supersymmetries as long as they have the right orientation. Thus
we see that the geometry displays the same supersymmetry
enhancement that we found in the field theory discussion of
section two. In conclusion, we see that the ${\cal N}=6$ $U(N)
\times U(N)$ superconformal Chern-Simons theory at levels $k$ and
$-k$ that we discussed in section two is the same as the
low-energy theory living on $N$ M2-branes probing a
$\bf{C}^4/\bf{Z}_k$ singularity in M-theory, as we conjectured in
the previous section.

\subsec{The special case of $k=1$}

Note that for the case $k=1$ we obtain M2-branes in ${\bf R}^8$.
Thus, the conformal Chern-Simons theories provide a (strongly
coupled) field theory description for the theory of M2-branes in
flat space. In this description the conformal symmetry is manifest
but not all the supersymmetries are manifest; we only see ${\cal
N}=6$ supersymmetry, and only an $SO(6)$ subgroup of the
R-symmetry is manifest.

In this special case we can also argue that the IR limit of the
Yang-Mills-Chern-Simons theory is the same as the IR limit of the
${\cal N}=8$ $U(N)$ SYM theory, by using the same arguments used for
deriving mirror symmetry between the IR limits of $2+1$
dimensional gauge theories realized in type IIB string theory
\refs{\IntriligatorEX,\HananyIE}, without lifting the
configuration to M-theory. Consider the S-dual of the ${\cal N}=3$
brane construction in \CSconfiguration. This gives $N$ D3-branes
on a circle, intersecting a D5-brane along 012345 and a $(k,1)$
fivebrane along $012[3,7]_\theta[4,8]_\theta[5,9]_\theta$. In
general, the low-energy theory on such branes in rather
complicated. However, for $k=1$ we can shift the axion by $2\pi$
and turn the $(1,1)$ fivebrane into an NS5-brane (without changing
its orientation). The low-energy
theory on the D3-branes is then a (single) $U(N)$ ${\cal N}=4$
supersymmetric gauge theory, coupled to an adjoint hypermultiplet
and a hypermultiplet in the fundamental representation (from the
D3-D5 strings). This is well-known to be equivalent at low
energies to the ${\cal N}=8$ SYM theory (without the fundamental
hypermultiplet). In the brane language this can be seen from the
fact that we obtain the same low-energy theory without the
NS5-brane, but then an additional S-duality gives $N$ D3-branes
intersecting a single NS5-brane which reduces to the ${\cal N}=8$
SYM theory. Since the low energy theories are independent of the
axio-dilaton, these arguments imply that the Chern-Simons-matter
theory described in the previous section with $k=1$, which arises
as the IR limit of our original configuration, has the same IR
limit as the $U(N)$ ${\cal N}=8$ SYM theory (which is equivalent
to the theory of $N$ M2-branes in flat space).

\subsec{ ${\cal N}=4$ supersymmetric brane configuration }

It is interesting to note that there is a brane configuration that is closely connected to
our previous discussion which preserves ${\cal N}=4$ supersymmetry. In the above
discussion we set the Ramond Ramond scalar, $\chi$,  to zero.
% and the string coupling to one.
If we consider more generic values of $\chi$, then in order to preserve ${\cal N}=3$
supersymmetry we need to rotate the $(1,k)$ fivebrane so as to maintain \relang .
It turns out that for particular values of the $\tau$ parameter of type IIB string theory
given by
\eqn\partval{
 \tau = { - k + i a k \over 1 + a^2 }
}
(for positive $a$)
the angle between the branes becomes $\theta = \pi/2$ so that they are orthogonal to each other.
At these particular values the supersymmetry is enhanced from ${\cal N}=3$ to ${\cal N}=4$.
This is shown in more detail using the M-theory lift in appendix B.
Changes in $\tau$ are not expected to affect the low energy field theory living on the branes.
Thus, we see that the IR theory living on this brane is still the conformal Chern-Simons theory
that we had above. One can see explicitly (see appendix B)
that these changes of the asymptotic moduli do
not change the structure of the ${\bf R^8/Z_k}$ singularity that we had in the M-theory
lift.

We do seem to have an apparent puzzle, however, since we do not expect to have three dimensional
Yang-Mills Chern-Simons theories with ${\cal N}=4$ supersymmetry. This puzzle is resolved by noticing that
for $\tau$ of the form \partval\ we cannot go to weak coupling in the bulk, without doing an S-duality.
 If we try to produce a small three dimensional gauge coupling on the brane by making the
size, $L$,
 of the direction 6 very large
  (recalling that ${ g_3^2} \sim { 1 \over {\rm Im}(\tau)  L}$), we find that there are
  Kaluza-Klein modes with a mass of the order of $1/L$ that we cannot decouple since we cannot
  make ${\rm Im}(\tau)$ large. Thus, we never obtain a weakly coupled $d=3$ ${\cal N}=4$ Yang-Mills
  Chern-Simons theory.

 \newsec{The dual gravitational backgrounds of M-theory and type IIA string theory}

The discussion of the previous sections suggests that the Chern-Simons-matter theories constructed
 in section 2 are dual to the conformal field theory living at low energies on $N$ M2-branes
 probing a ${\bf C}^4/{\bf Z}_k$ singularity. This theory has a dual gravitational
 description in terms of M-theory on $AdS_4\times S^7/{\bf Z}_k$. In this section we discuss various aspects of this gravity dual.

\subsec{Supergravity backgrounds}

% We have argued that the conformal field theory lives on $M2$ branes probing an $R^8/Z_k$
% singularity. If we think of  $R^8$ as four copies of the complex plane $R^8 = C^4$, then
If we write the transverse space to the M2-branes using four complex coordinates $z_i$
($i=1,2,3,4$), then
 the ${\bf Z}_k$ quotient we are considering acts as
 \eqn\qoutcon{
 z_i \to e^{ i { 2 \pi \over k } } z_i.
 }
  The gravity dual of $N$ M2-branes in flat space is  $AdS_4 \times S^7$, and
 we simply need to quotient this by this ${\bf Z}_k$. Since the ${\bf Z}_k$ preserves
 an $SU(4)\times U(1)$ isometry symmetry (acting on the $z_i$ in an obvious way), it is
 natural to use the description of $S^7$ as an $S^1$
 fibration over ${\bf CP}^3$. In this Hopf fibration the circle has a constant radius, and the ${\bf Z}_k$ quotient is simply making
 this radius smaller \nilssonpope .
 More explicitly, we begin with the $AdS_4\times S^7$ solution of eleven dimensional
 supergravity with $N'$ units of four-form flux (we will relate this to $N$ and $k$ later),
 which has the metric and four-form
\eqn\elevnd{ \eqalign{
ds^2 =&  { R^2 \over 4 } ds^2_{AdS_4} + R^2 ds^2_{S^7},
\cr
F_4 \sim  & N' \epsilon_4,
\cr
R/l_p =  & ( 2^5 \pi^2 N')^{1/6},
}}
where the metrics $ds^2_{AdS_4}$ and $ds^2_{S^7}$ have unit radius.
We can write the metric of $S^7$ as
\eqn\metrise{ \eqalign{
ds^2_{S^7} = & ( d \varphi' + \omega)^2 + ds^2_{CP^3},
 \cr
   ds^2_{CP^3} =& { \sum_i d z_i d \bar z_i  \over \rho^2 } -
    { | \sum_i z_i d \bar z_i |^2 \over \rho^4} ~,~~~~
 \rho^2 \equiv \sum_{i=1}^4 |z_i|^2,
 \cr
 d \varphi' + \omega  \equiv & { i \over 2  \rho^2 } (  z_i d \bar z_i - \bar z_i d z _i ),
 \cr
 d\omega = & J = { i}   d \left({ z_i \over \rho}\right)  d \left({ \bar z_i \over \rho }\right),
  }}
 where $\varphi'$ is periodic with period $2\pi$ and $J$ is proportional   to the
  K\"ahler form on $\bf{CP}^3$.   We now perform
 the $\bf{Z}_k$ quotient. For that purpose we write $\varphi' = \varphi/k$,
 with $\varphi \sim \varphi + 2 \pi$.
  The metric then becomes
 \eqn\nermetr{
 ds^2_{S^7/{\bf Z}_k} = { 1 \over k^2 } ( d \varphi + k \omega)^2 + ds^2_{CP^3}.
 }
 Since the volume of this space is smaller by a factor of $k$ than the original volume,
 in order to have a properly quantized flux on the quotient space we need that
 $N' = k N$ where $N$ is the final number of flux quanta on the quotient space.

 The spectrum of supergravity fields on this background (which are the chiral
 primaries that are visible in the gravity approximation) is simply the projection of the
 original spectrum on $AdS_4\times S^7$ \BiranIY\ onto the ${\bf Z}_k$-invariant
 states \refs{\nilssonpope,\HalyoPN}\foot{ See \nilssonpope\ for a nice table of the $U(1)$-invariant
 states. }. The lowest components of the chiral primary multiplets are
 \AharonyRM\ in $l$'th symmetric traceless products ($l=2,3,\cdots$) of the ${\bf 8}_v$
 representation of $SO(8)$, which decomposes as ${\bf 4}_1 + {\bf \bar 4}_{-1}$ under
 $SU(4)\times U(1)$. It is easy to see that these precisely match the spectrum of
 chiral primaries we discussed in section 2, when we identify the $U(1)$ isometry
 with $U(1)_b$. We have states that
 carry no
 $U(1)_b$ charge (and are thus preserved by the orbifold projection for any $k$) and
 also states that carry a $U(1)_b$ charge which is a multiple of $k$.\foot{
 In the special case of $k=1$, the field theory
 spectrum contains also the   state with $l=1$ describing the center of mass motion of
 the branes. In the gravity description it is
 a ``singleton'' field which can be gauged away to the boundary and
  is decoupled from the bulk physics.}

The radius of the ${\bf CP}^3$ factor is large whenever $N'=Nk \gg 1$. However,
the radius of the $\varphi$ circle in Planck units is of the order of
$R/kl_p \propto (Nk)^{1/6} / k$. Thus, the M-theory description is valid whenever
$k^5 \ll N$, and when $k$ increases the circle becomes small and we can reduce to
a weakly coupled type IIA string theory using the usual formulas.

 The reduction to type IIA gives the string frame metric, dilaton and Ramond-Ramond forms (setting $\alpha'=1$) \refs{\WatamuraHJ,\nilssonpope}
 \eqn\twoasol{ \eqalign{
 ds^2_{string} = & { R^3 \over k} ( { 1 \over 4 } ds^2_{AdS_4} + ds^2_{CP^3 } ),
 \cr
 e^{2 \phi} = & { R^3 \over k^3 } \sim { N^{1/2} \over k^{5/2} }= { 1 \over N^2 } \left( {
 N \over k } \right)^{5/2},
 \cr
 F_{4} = & { 3 \over 8 }  {  R^3} \hat \epsilon_4 ,
 \cr
 F_2 = & k d \omega = k J,
 }}
 where $\hat \epsilon_4$ is the epsilon symbol in a unit radius $AdS_4$ space.
 We see that we have an $AdS_4 \times {\bf CP}^3$ compactification
 of type IIA string theory (supergravity) with $N$ units of $F_4$ flux on $AdS_4$ and $k$ units of $F_2$ flux on the ${\bf CP}^1 \subset {\bf CP}^3$ 2-cycle.

The radius of curvature in string units is
\eqn\radcurv{
 R^2_{str} =   { R^3 \over k} =  2^{5/2} \pi  \sqrt{ N \over k }  =  2^{5/2} \pi \sqrt{ \lambda}~,
 % ~~~~~~~
 %\lambda \equiv { N \over k }
 }
 where, as before, $\lambda \equiv N/k$ is the 't Hooft coupling.
 It is interesting that the functional dependence of the curvature on the 't Hooft coupling
 is the same as in $d=4$ ${\cal N}=4$ SYM \MaldacenaRE. As expected, for fixed 't Hooft coupling the
 string coupling goes like $1/N$. Notice that the existence of a weakly coupled  string
 dual
  was guaranteed by the fact that the field theory had a (discretely) adjustable parameter
 that enables us to go to weak coupling for fixed $N$. Therefore one can
 take the usual
 't Hooft limit, which in this case is $k, N \to \infty $ with $N/k$ fixed.
 Of course,   when $\lambda $ becomes of order one, the approximation of type IIA string
 theory by supergravity (which we used to write the solution above) breaks down,
 since the curvature becomes of order the string scale.
  The field theory is weakly coupled when $\lambda \ll 1$.
  The type IIA supergravity approximation is valid in the regime where
  \eqn\refim{
  1 \ll \lambda ~,~~~~~~~~~~{ N^{1/2} \over k^{5/2} } = {\lambda^{5/2} \over N^2 } \ll 1 ~,~~~~~~
  }
  while the eleven dimensional supergravity approximation is valid when $N \gg k^5$.

As in other cases of the AdS/CFT correspondence \WittenZW, the finite temperature behavior
is governed in the gravitational description by an AdS black hole.
 The finite temperature partition function (in volume $V_2$ and temperature $T$) thus has a behavior which is very similar to the one for M2-branes
 \klebanovtseytlin,
 \eqn\fintpa{
  \beta F  = - 2^{7/2} 3^{-2} \pi^2 { ( N k )^{3/2} \over k }  V_2 T^2  =
    - 2^{7/2} 3^{-2} \pi^2
  N^2 { 1 \over \sqrt{ \lambda} } V_2 T^2.
  }
  This expression is valid for $\lambda \gg 1$ (the same expression arises both from
  eleven dimensional supergravity and from type IIA supergravity), and it receives corrections
  going (in the large $N$ limit) as inverse powers of $1/\sqrt{\lambda}$.
  Notice that it has both the characteristic $N^{3/2}$ behavior of M2-branes and
  the expected $N^2$ behavior of large $N$ gauge theories for fixed $\lambda$. Of course,
  it would be very nice to derive \fintpa\ directly from the field theory point of view.
  At weak coupling ($k \gg N$) the field theory has a free energy given by the free field theory
  result
  \eqn\freeft{
  \beta F \sim -N^2  V_2 T^2  \left[  { 7 \zeta(3) \over {\pi} } + O(\lambda) \right] ~,~~~~~~~\lambda \ll 1,}
  with corrections going as
  integer powers of $\lambda$. We see that as we go to strong
  't Hooft coupling the entropy decreases as $1/\sqrt{\lambda}$ giving rise to \fintpa.
  This is different from the behavior of $d=4$ ${\cal N}=4$ SYM, where the entropy goes
  to a constant in the strong coupling limit \GubserDE .

 It is interesting to understand the symmetries of the gravitational solutions. The M-theory
 solution clearly has an $SU(4) \times U(1)$ symmetry. The supercharges are  in the
 ${\bf 6}_0$ representation of this group \refs{\nilssonpope,\susywosusy}, so $SU(4)$ is
 the R-symmetry group and the $U(1)$ is a global symmetry, corresponding to shifts in $\varphi$ in \nermetr . In the type IIA picture the $SU(4)$ symmetry remains as a geometric symmetry of ${\bf CP}^3$.
 We are tempted to identify the $U(1)$ symmetry of M-theory with the RR symmetry in type IIA
 under which the D0-branes are charged. To a first approximation this is correct. However,
 we should remember that in an M-theory background with fluxes these symmetries also involve shifts
 in certain background potentials \BiranIY.
 In the type IIA picture we might naively think that we have two $U(1)$ gauge fields -- the
 RR 1-form potential, and the 3-form potential integrated over ${\bf CP}^1 \subset {\bf CP}^3$, whose electric and magnetic charges are carried by
 D0-branes, D2-branes wrapped on ${\bf CP}^1 \subset {\bf CP}^3$, D4-branes
 wrapped on ${\bf CP}^2 \subset {\bf CP}^3$ and D6-branes wrapped on ${\bf CP}^3$. We will see that in the presence of the $F_2$ flux one combination of these gauge fields is
 Higgsed and becomes massive. The massless one corresponds to  the symmetry generated by
 \eqn\jgen{
 J = k Q_0 + { N } Q_4,
 }
 where $Q_0$ and $Q_4$ are the D0-brane and wrapped D4-brane charges, respectively.
 One way to see this is from the fact that
 a maximal giant graviton M5-brane \McGreevyCW\ (wrapped on $S^5 \subset S^7$)
 has charge $J  = N' = N k$ in the covering space.
 After Kaluza-Klein reduction this becomes a D4-brane wrapping the ${\bf CP}^2$.
 %The ${\bf Z_k}$ quotient requires $J$
 %to have values which are multiples of $k$.
  We identify the current \jgen\ with the baryon number
 current $J_b$ in the field theory discussion.

From the M-theory point of view it is clear that there is only one massless $U(1)$ gauge
field since the only massless gauge fields before doing the orbifold are the $SO(8)$ ones, and
the orbifold leaves only one $U(1)$ (and the $SU(4)$, of course). It is instructive
to see how this comes about from the type IIA point of view.
Let us write down the equations of motion
involving the RR fluxes and the NS-NS 3-form flux:
 \eqn\fluxeqm{ \eqalign{
 &  d \tilde F_4 = - F_2 \wedge H_3 ~,~~~~~~~d * \tilde F_4 = \tilde F_4 \wedge H_3  ~,~~~~~~~
  d H_3 =0 ~,~~~~d F_2 =0~, \cr
  & d * ( e^{-2 \phi}  H_3 ) = - F_2 \wedge *  \tilde F_4 + { 1 \over 2}
   \tilde F_4 \wedge \tilde F_4 ~,~~~~~~ d * F_2 = H\wedge * \tilde F_4~,
 \cr
 & \tilde F_4 = dA_3 - A_1\wedge H_3 ~.
% =
%  d \hat A_3 - F_2 \wedge B_2 ~,~~~~~~~~\hat A_3 = A_3 + A_1 \wedge B_2.
 }}

Writing down an effective field theory on $AdS_4$, we naively obtain two $U(1)$ field
strengths :
$F^{D0}$ containing the $AdS_4$ components of $F_2$, and ${\tilde F}^{D2}$ containing the
$AdS_4$ components of the 2-form $\int_{{\bf CP}^1} {\tilde F}_4$.
We also need to keep the $H_3$ field in the $AdS_4$ directions. This
will be dual to an axion. Keeping only these fields we then see that the equations
\fluxeqm\ in the background \twoasol\ imply the four dimensional equations:
  \eqn\cadf{\eqalign{
   & d F^{D0} =0 ~,~~~~ *_4 d *_4 F^{D0} = N *_4 H_3 ~,~~~~~~ *_4 d *_4 \tilde F^{D2} = 0 ~,~~~~~~
   d \tilde F^{D2} = k H_3 ~,~~~
   \cr
   & d *_4 ( e^{-2 \phi} H_3) = N F^{D0}   - k *_4 \tilde F^{D2}~.
   }}
   We now define $ F^{D4} = *_4 \tilde F^{D2} $, and the definition of the axion $a$ dual
   to the $B$ field is deformed to
   \eqn\actica{
    da = * e^{-2 \phi} H_3 - N A^{D0} + k A^{D4}~.
    }
    Then, the equations for the four dimensional fields become
    \eqn\newequ{ \eqalign{
     & d F^{D0} =0 ~,~~~~~~d F^{D4}  =0 ~,~~~~~~~~
     \cr  &
      *_4 d *_4   F^{D0 } = N e^{2\phi} ( da + N A^{D0} - k A^{D4} )
     ~,~~ *_4 d*_4 F^{D4 } = - k e^{2\phi} ( da + N A^{D0} - k A^{D4} )~.
   }}
   Thus we see that one linear combination of these fields becomes massive (by swallowing
   the axion field) while the other
   remains massless. The combination that remains massless can be parameterized in terms
   of a gauge field $A^J$ as $A^{D0} = kA^J$, $A^{D4} = {N  } A^J$, which is the
   same as the statement in \jgen.

   We should also note that $k$ units of D4-brane charge can be turned into $N$ D0-branes
   by an NS5-brane instanton. This is easiest to see in the covering space as the statement that
   $N'=Nk$ units of momentum ($N$ units of D0-brane charge) can become a maximal giant
   graviton M5-brane \McGreevyCW, which in the quotient space maps to $k$ D4-branes wrapping ${\bf CP}^2$.

\subsec{Particle-like branes}

   Let us analyze more explicitly the various particle-like branes in our background.
   In the M-theory description, $S^7/{\bf Z}_k$ does not have any integer-valued homology
   classes, but it has non-trivial ${\bf Z}_k$ homology classes corresponding to the circle
   and to the product of the circle with ${\bf CP}^1$ or ${\bf CP}^2$ inside ${\bf CP}^3$.
   We can thus get non-trivial particle-like branes only by wrapping an M5-brane on the circle
   times ${\bf CP}^2$, and non-trivial string-like
   % OA - minor correction
   branes
   %brane
   only by wrapping an M2-brane
   on the circle (leading to the type IIA fundamental string).

   In the type IIA description, the D0-brane carries $k$ units of charge under $J$ (it carries $k$
   units of momentum on the covering space,
    since the circle in the orbifold is smaller by a factor of $k$). Thus, it is natural to identify it with the operators in the field theory of the form $C^k$
   which were noted in the field theory discussion. The D0-brane is moving in a magnetic field
   on the ${\bf CP}^3$. Thus we expect to have a large number of states, proportional to
   $ { 1 \over 6}  \int ( { F_2 \over 2 \pi } )^3 = { k^3 \over 6 }
   \int { J^3 \over ( 2 \pi)^3 } = { k^3 \over 6 }  $.
   This is indeed the scaling in $k$ (at large $k$)  for the
    dimension of a $k$'th symmetric product of ${\bf 4}$'s of $SU(4)$. This gives correctly the
    chiral primaries with $J$ charge equal to $k$ and lowest energy. In addition we have other states with the same $U(1)$ charge and higher $SU(4)$ representations.
    In the M-theory description they correspond to
    spherical harmonics with angular momentum $n > k $ but only $k $ units of $U(1)$ charge in the covering space. These can also be mapped to chiral operators, as discussed in
    section 2.
    %It would be interesting to perform a precise matching of the D0-brane states
    %in type IIA string theory with the field theory.
%    We have not matched these in detail to the field theory operators.

   The D2-branes wrapped on ${\bf CP}^1$ in ${\bf CP}^3$ have a worldvolume
   coupling of the form $\int A \wedge F_2 \sim k \int dt A \wedge J  $, where $A$ is the worldvolume
   $U(1)$ gauge field. This implies that
    they can only exist if $k$ strings end on them. Thus we see we cannot have
   an isolated D2-brane, consistent with the fact that this would carry a magnetic charge
   under a $U(1)$ field that was Higgsed.
   On the other hand, we can have a D2-brane with $k$ strings ending on it. This is   related to the fact
   that Wilson lines with $k$ fundamentals under one gauge group can be screened, since as
   usual in AdS/CFT \MaldacenaIM\ in the 't Hooft limit we can identify the Wilson lines
   with fundamental strings in the
   bulk.
   % JM
   %In the bulk
   %this screening corresponds to the D2-brane moving to the boundary, and completely screening the
   %Wilson line. Namely, at the D2-brane the $k$ lines are combining into an
   % unobservable Wilson line in the $k$'th symmetric
   %representation.
   In the gauge theory this is related to the fact that adding 't Hooft or monopole fluxes can screen
   the $SU(N)$ quantum numbers.
   % JM
   The M-theory lift of a configuration with $k$ strings ending on a D2-brane is
   simple. As an example we can take $k$ M2-branes that are extended in the time
   direction and along the $z_1$ plane. They sit at $z_2 = ({\rm const}) e^{ i 2 \pi n/k}$ for
   $n=0,1,\cdots, k-1$ and at $z_3=z_4=0$. Then, for large $|z|$ we see that we can neglect the
   non-zero value of $z_2$ and we basically have a brane that wraps the $U(1)$ fibration. Since
   we have $k$ M2-branes the total string charge is $k$. Near $z_2 \sim ({\rm const} )$ we should
   think of the brane as an M2-brane transverse to the circle fibration, which can be interpreted
   as a D2-brane.

   The D4-brane wrapped on ${\bf CP}^2$ carries $N$ units of the massless $U(1)$ charge
   $J$ which we identified with the baryon number charge.
   From the M-theory point of view this starts its life as a maximal giant graviton M5-brane
   wrapping an $S^5$ in $S^7$. After the ${\bf Z_k}$ quotient
   this is a brane wrapping a non-trivial
   ${\bf Z}_k$ homology cycle that consists of the fiber direction and a
   ${\bf CP^2 } \subset {\bf CP^3 }$.
   %
   %This object
   %only exists when $N/k$ is an integer, or when we have $n$ D4-branes with $n N/k$ an integer,
   %otherwise it is projected out by the orbifold. Notice that   this
   %implies that the generator of the orbifold is the full conserved
   %charge $J$ with D0 and D4 contributions as in
   %\jgen. $J$ has to be a multiple of $k$.
   We can identify these wrapped branes with the di-baryon  operators discussed in the field
   theory section.
   % though there is no distinction between these operators and the operators
   %dual to D0-branes (one can think of the wrapped D4-brane as a ``giant graviton'' version
   %of many D0-branes).

   Finally, the D6-brane wrapped on ${\bf CP}^3$ has a coupling on its worldvolume involving
   $\int A \wedge *F_4 $ which implies that $N$ strings should end on it. This is the
   baryon vertex, as in \wittenbaryon .

   Notice that the presence of the D4-brane and the D6-brane suggests that we are not
   dealing precisely with the $U(N)\times U(N)$ theory, since in this theory these objects
   would only be allowed when $N$ is a multiple of $k$. Rather, the theory we find resembles
   the $(SU(N) \times SU(N)) / {\bf Z}_N$ theory, in which this constraint is not present.
   Presumably, the difference between the $U(N)\times U(N)$ theory and the theory that we find
   lies (as in other examples of the AdS/CFT
   correspondence) in the behavior of the subtle
   ``singleton'' modes that live on the boundary of $AdS$ space \MaldacenaSS, since
   the bulk physics arising from a brane configuration with $U(N)$ gauge symmetry typically captures only the $SU(N)$ dynamics (see, e.g. \AharonyQU). It would be nice to
   elucidate this point further.

Note that the wrapped D2-brane and the wrapped D6-brane both carry a magnetic charge
under the massless $U(1)$ field \jgen.  They also carry magnetic charge under the $U(1)$ that
was Higgsed. The latter is responsible for the presence of strings that end on the object.
We can form a combination of D6 and D2-branes that carries no charge under the Higgsed $U(1)$.
In our conventions, we see from \newequ\ that a D6-brane has $N$ strings ending on it whearas
a D2-brane has $k$ strings leaving it. Thus if we consider a system of $k$ D6-branes together
with $N$ D2-branes then we find that they form an object with no magnetic charge under the
Higgsed $U(1)$, so that there need not be any strings ending on it. This looks like a perfectly
reasonable localized excitation in the bulk. This excitation, however, carries magnetic
charge under the massless $U(1)$ gauge field.
The boundary conditions for a $U(1)$ gauge field
on $AdS_4$ only allow for either electric or magnetic charges \wittensl, and in our case
the allowed charges are electric, so these  branes are not allowed. As explained in
\wittensl, one can change the boundary conditions and the dual field theory so that they
would be allowed (and the electrically-charged branes would not be allowed).

 \subsec{High spin operators}

 A rotating string as in \gkp\ is dual to a high spin operator. We can compute its anomalous
 dimension in the large $\lambda$ limit. The result is the same as in \gkp\ when it is expressed in
 terms of the radius of $AdS$ in string units.
 In
 our case  we have
 \eqn\firtw{
 { R^2_{AdS_4} \over \alpha'  } =  { 2^{1\over 2} \pi \sqrt{ \lambda} } =
  2^{1\over 2} \pi  \sqrt{ N \over k }.
 }
 Thus, we find that the anomalous dimension of high spin operators goes like
 \eqn\finequ{
 \Delta - S = f(\lambda) \log S  , ~~~~~ f(\lambda ) = { R_{AdS_4}^2 \over \pi \alpha' } =
  { \sqrt{2 \lambda }  }  ~,~~~~
 \lambda \gg 1.
 }
 where $f(\lambda)$ is also the cusp anomalous dimension.
 As shown in \aldayjmtwo, the logarithmic behavior is a general property of any conformal field
 theory with gauge fields in any number of dimensions.
 The weak coupling computation was done for general theories in
  \gaiottoyin \foot{
 In \gaiottoyin\ the general computation was performed for an ${\cal N}=2$ theory
 with $N_f$ flavors. The leading order term does not depend on the matter interactions.
 Thus we obtain our result by taking $N_f = 4 N $ and multiplying the result in \gaiottoyin\
  by a factor
 of two, because in \gaiottoyin\ the matter was in the fundamental, where now it is in the
 bifundamental. In other words, in \gaiottoyin\ they considered an open string while here
 we are considering a closed string.}
 \eqn\weakcup{
  f(\lambda) = \lambda^2 + O( \lambda^3) ~,~~~~~~\lambda \ll 1.
  }

  Of course, it is natural to wonder if this string theory is integrable and whether
  there is a formula for the cusp anomalous dimension for all $\lambda$, as in
  %\BeisertEZ
\ref\BeisertEZ{
  N.~Beisert, B.~Eden and M.~Staudacher,
  ``Transcendentality and crossing,''
  J.\ Stat.\ Mech.\  {\bf 0701}, P021 (2007)
  [arXiv:hep-th/0610251].
  %%CITATION = JSTAT,0701,P021;%%
}, despite the fact that in our case $\lambda$ is restricted to rational values.

\newsec{Conclusions}

In this paper we constructed and discussed $U(N)\times U(N)$ Chern-Simons-matter theories
with ${\cal N}=6$ supersymmetry. We argued that these theories were equivalent to the low-energy
 theory on $N$ M2-branes at a ${\bf C}^4/{\bf Z}_k$ singularity. For $k \gg N$ these theories are
  weakly coupled, and we
argued that for $N \gg 1$ and $k \ll N$ they had dual
% OA - minor addition
weakly curved
gravitational descriptions in term of M-theory (when $k \ll
N^{1/5}$) or type IIA string theory (when $N^{1/5} \ll k \ll N$).
The string theory background describes the theory in the 't Hooft
large $N$ limit. It would be interesting to understand whether the type IIA
string theory is integrable or not.

Since Chern-Simons theories play an important role in condensed
matter systems,
 it is interesting to ask whether there are any other Chern-Simons-matter theories which have
a dual gravitational description. The gravity description means that they
can be effectively solved at strong coupling. Having such a dual
gravitational description predicts many properties of these
theories, and might give insight on certain condensed matter
systems.

One of the main lessons we can draw   is that performing a ${\bf Z}_k$ orbifold has
allowed us to find an explicit Lagrangian description of interesting M2-brane theories (which
becomes weakly coupled for large $k$). It seems
natural to attempt to repeat this procedure in spaces which are not locally ${\bf R}^8$ to obtain
a  variety of  theories. Hopefully this can eventually shed some light on the field theory duals
of $AdS_4$ compactifications in the string landscape.

The case $k=1$ is particularly interesting
 since  our theories are believed
to be equivalent to the ${\cal N}=8$ SCFT living on $N$  M2-branes in flat space. We provide
an explicit Lagrangian for this SCFT. This Lagrangian displays explicitly only
an  ${\cal N}=6$ supersymmetry. The spectrum of protected operators in our theory becomes
$SO(8)$ invariant for $k=1,2$. The $SO(8)$ generators which are not in $SO(6)$
 seem to be non-locally
realized in terms of the fields in our Lagrangian.
In particular, some of the
$SO(8)_R$ generators can
transform a field $C$
into $C^\dagger$. Thus, it seems that such generators should involve Wilson lines in the symmetric
product of two fundamentals of the first gauge group and two anti-fundamentals of the second.
Interestingly, these Wilson lines are unobservable precisely for $k=1,2$, which are the
cases where we have $SO(8)$ symmetry. So, in these cases the generators may be local
operators. It is an interesting problem  to construct these
generators explicitly. This is analogous  to the construction of the supercharges for
a compact boson in two dimensions which is at the special radius where it is supersymmetric.

Of course, for $k=1,2$ our theory is
very strongly coupled, so it is not clear what we can learn about
it from our description, except for properties which are protected by supersymmetry. Many
of these properties, such as the spectrum of chiral operators, are already known, either from
the description of these theories as the IR limit of the ${\cal N}=8$ $SU(N)$ SYM theory or
from their description as M-theory on $AdS_4\times S^7$, and we can verify these results
using our description.
%of these theories at large $N$, and there is a natural generalization
%to finite $N$ (involving the truncation of the spectrum of independent chiral operators coming
%from the product of $l$ fundamentals of $SO(8)$ when $l=N$); in our formalism we can %explicitly
%verify this.
It is not clear yet how to use our formalism to compute things
which are not protected by supersymmetry, such as the entropy of
the theory at finite temperature. In principle it might be
possible to compute this by putting our theory on a lattice. Of
course one could also do such a computation   using the
alternative definition of these SCFTs as the IR limit of the
${\cal N}=8$ $U(N)$ SYM theory. In our formalism the conformal
symmetry is manifest, but not all the supersymmetries are
manifest, whereas if we view the theory as an IR limit of ${\cal
N}=8$ super Yang-Mills the supersymmetries are manifest but the
conformal symmetry is not. For other proposals for  a
theory of multiple M2-branes in flat space, see \multiplemtwo .

Another interesting problem is  the generalization of our discussion to the
case of the $U(N) \times U(M)$ ($SU(N)\times SU(M)$) gauge theory. The generalization
of our field theory discussion is straightforward, and these theories still have
${\cal N}=6$ superconformal symmetry. In the gravitational dual description, it
seems that these theories may be related to turning
on $n=N-M$ units of
 ${ \bf Z_k }$ torsion  $G_4$ flux in M-theory, if  $|N-M| \leq k$.
 % JM
 Theories where $|N-M|>k$ do not appear to exist as superconformal theories.
 Notice that in such theories
 at least one of the gauge groups would be strongly coupled.

It would also be interesting to understand how to obtain the
$SU(N) \times SU(N)$ theories directly from  a string construction. One can consider
putting the Klebanov-Witten theory \KlebanovHH\ on a compact
circle and deforming it by the appropriately fine tuned three dimensional supersymmetric
Chern-Simons term. At the level of supergravity this
could lead to a flow between a space which is locally $AdS_5 \times T^{1,1}$ and
the $AdS_4 \times S^7/{\bf Z}_k$ geometry.

It seems that our theories should also admit massive deformations preserving all the
supersymmetries, similar to the ones considered in \BenaZB .

\vskip 1cm

  {\centerline {\bf Acknowledgements }}

We would like to thank N. Arkani-Hamed, J. Bagger,  D. Gaiotto, N. Itzhaki, I. Klebanov,
 N. Lambert, G. Moore, S. Thomas, E. Witten and X. Yin for discussions.
OA would like to thank the Institute for Advanced Study and the
University of Pennsylvania for hospitality. The work of OA was
supported in part by the Israel-U.S. Binational Science
Foundation, by a center of excellence supported by the Israel
Science Foundation (grant number 1468/06), by a grant (DIP H52) of
the German Israel Project Cooperation, by the European network
MRTN-CT-2004-512194, and by Minerva. OB gratefully acknowledges
support from the Institute for Advanced Study. OB also thanks the
Institute for Nuclear Theory at the University of Washington for
its hospitality and the US Department of Energy for partial
support during the completion of this work. The work of OB was
supported in part by the Israel Science Foundation under grant
no.~568/05. The work of DJ was supported in part by DOE grant
DE-FG02-96ER40949. This work was supported in part by DOE grant
\#DE-FG02-90ER40542.

\appendix{A}{Explicit verifications of symmetries}

\subsec{Explicit verification of $SU(4)_R$ symmetry of the bosonic potential}

In this subsection we compute explicitly the bosonic potential and we show that it is
invariant under $SU(4)_R$.

 We consider the $U(N) \times U(N)$ theory.
 We have
 \eqn\computsup{
 %\eqalign{
 W = {2 \pi \over k}  \epsilon^{a b} \epsilon^{\dot a \dot b}
 \Tr[ A_a B_{\dot a } A_{b} B_{\dot b} ]~,
% \cr
\quad
 { \partial W\over \partial A_a } = {4\pi \over k} \epsilon_{a b} \epsilon_{\dot a \dot b}
   B_{\dot a } A_{b} B_{\dot b}~,
%   \cr
\quad
   { \partial W\over \partial B_{\dot a} } = {4\pi \over k} \epsilon_{a b} \epsilon_{\dot a \dot b}
  A_b B_{\dot b } A_{a}~,}
%  ,\cr}}
%
so that the scalar potential arising from the superpotential is
\eqn\vsup{V_{sup} = |\partial W|^2 =
{16 \pi^2 \over k^2} \left(\epsilon_{\dot a \dot b } \epsilon_{\dot c \dot d } \Tr[ B^\dagger_{\dot b} A^\dagger_a
  B^\dagger_{\dot a } B_{\dot c} A_a B_{\dot d} ]
  + \epsilon_{ab}\epsilon_{cd} \Tr[ A^\dagger_b B^\dagger_{\dot a } A^\dagger_a A_c B_{\dot a }
  A_d ]\right).
  }

  In addition we find that
  \eqn\find{ \eqalign{
  {k\over {2\pi}} \sigma_{(1)} = &  A_a A^\dagger_a - B^\dagger_{\dot b}  B_{\dot b}~,
  \cr
 - {k\over {2\pi}} \sigma_{(2)} = &  B_{\dot a }B^\dagger_{\dot a } - A^\dagger_a A_a   ~~~
 \to ~~~
 {k\over {2\pi}} \sigma_{(2)} =  A^\dagger_a A_a  -  B_{\dot a }  B^\dagger_{\dot a }~,
 }}
where
 the Chern-Simons piece of the potential takes the form
 \eqn\chersimpot{
 V_{CS} = \Tr[ A_c  A^\dagger_c \sigma_{(1)}^2- 2 A^\dagger_c \sigma_{(1)} A_c \sigma_{(2)} + A^\dagger_c A_c
 \sigma_{(2)}^2 ] + \Tr[ B_{\dot c}  B^\dagger_{\dot c} \sigma_{(2)}^2 -
 2 B^\dagger_{\dot c } \sigma_{(2)} B_{\dot c }  \sigma_{(1)} + B^\dagger_{\dot c } B_{\dot c }  \sigma_{(1)}^2 ].
 }

 We would now like to write the sum of these two terms
 in an $SU(4)_R$-invariant way, using $C_I \equiv (A_1,A_2, B^\dagger_1, B^\dagger_{2} ) $.
 A bosonic potential which is $SU(4)$ invariant and has the structure of the ones above
 must be of the form
 \eqn\formpot{
 V \sim \Tr[ C_{I_1} { C^\dagger}^{J_1}  C_{I_2} { C^\dagger}^{J_2}  C_{I_3} { C^\dagger}^{J_3} ]
 }
where the indices need to be contracted in some way.
In general, there are four possible contractions (up to cyclic permutations).
The most general $SU(4)$-invariant potential is
\eqn\mostgenp{ \eqalign{
V = {4 \pi^2 \over k^2} \big( & a_1 \Tr[ C_{I_1} { C^\dagger}^{I_1}  C_{I_2} { C^\dagger}^{I_2}  C_{I_3} { C^\dagger}^{I_3} ] +
a_2  \Tr[ C_{I_1} { C^\dagger}^{I_2}  C_{I_2} { C^\dagger}^{I_3}  C_{I_3} { C^\dagger}^{I_1} ] +
\cr &
a_3  \Tr[ C_{I_1} { C^\dagger}^{I_1}  C_{I_2} { C^\dagger}^{I_3}  C_{I_3} { C^\dagger}^{I_2} ]
+ a_4   \Tr[C_{I_1} { C^\dagger}^{I_3}  C_{I_2} { C^\dagger}^{I_1}  C_{I_3} { C^\dagger}^{I_2} ] \big) .
}}
The coefficients can be determined by matching to the potential we wrote above in simple special cases. We can consider diagonal $C$'s and
demand that $V$ vanishes, we can set to zero some of the components of $C$, etc. In the end we obtain
$
3 a_1 = 3 a_2 = -1 ,  ~a_3 =2 ,~3 a_4 =-4 $.

In summary, we can write the potential as
\eqn\claimfin{ \eqalign{
V = & V_{sup} + V_{CS} =
\cr
= & {4 \pi^2 \over k^2} \big(  -{  1 \over 3}
 \Tr[ C_{I_1} { C^\dagger}^{I_1}  C_{I_2} { C^\dagger}^{I_2}  C_{I_3} { C^\dagger}^{I_3} ]
- { 1 \over 3 }   \Tr[ C_{I_1} { C^\dagger}^{I_2}  C_{I_2} { C^\dagger}^{I_3}  C_{I_3} { C^\dagger}^{I_1} ] +
\cr & ~~~~~
+ 2   \Tr[ C_{I_1} { C^\dagger}^{I_1}  C_{I_2} { C^\dagger}^{I_3}  C_{I_3} { C^\dagger}^{I_2} ]
 - { 4 \over 3 } \Tr[C_{I_1} { C^\dagger}^{I_3}  C_{I_2} { C^\dagger}^{I_1}  C_{I_3} { C^\dagger}^{I_2} ]
 \big) =
  %. }}
%I have checked with mathematica that these are the correct coefficients.
%
%Equivalently, we can  rewrite this potential as
%\eqn\rewrpot{
%\eqalign{
%V =& - 2 Tr( C_{i_1} { C^\dagger}^{[i_1}  C_{i_2} { C^\dagger}^{i_2}  C_{i_3} { C^\dagger}^{i_3 ]} ) +
%\cr
% &+   Tr( C_{i_1} { C^\dagger}^{i_1}  C_{i_2} { C^\dagger}^{i_3}  C_{i_3} { C^\dagger}^{i_2} ) -
%  Tr(C_{i_1} { C^\dagger}^{i_3}  C_{i_2} { C^\dagger}^{i_1}  C_{i_3} { C^\dagger}^{i_2} )
%\cr
\cr
= &  - {8 \pi^2 \over k^2} \Tr( C_{I_1} { C^\dagger}^{[I_1}  C_{I_2} { C^\dagger}^{I_2}  C_{I_3} { C^\dagger}^{I_3 ]} )
 +  {8 \pi^2 \over k^2} \Tr( C_{ I_1} { C^\dagger}^{[I_1}  C_{I_2} { C^\dagger}^{I_3]}  C_{I_3} { C^\dagger}^{I_2} ),
 }}
where the brackets $[~]$ mean antisymmetrization with a factor of 1/6 or 1/2 respectively.
We see that it is indeed invariant under the $SU(4)$ symmetry.
% But this is not
%too illuminating. Perhaps there is a more illuminating way to write it. One could also study
%the space of solutions of $V=0$ (the moduli space).

\subsec{Enhanced flavor symmetry for gauge group $SU(2)\times SU(2)$}

\def\a{\alpha}
\def\b{\beta}

In this appendix we verify that in the case of $SU(2)\times SU(2)$ the superpotential \nwnsix\
is invariant under an enhanced $SU(4)$ flavor symmetry rotating the chiral multiplets.
%  We would now like to ask whether the theory has more symmetries in the case that
%  the gauge group is $SU(2)$
In this case we can also think of the fields $B_i$ as bifundamentals, rather
than anti-bifundamentals; more precisely we can write
    \eqn\newdef{
     (\hat B_i)_{\a}^{~\b} = \epsilon_{\a \a'} \epsilon^{\b \b'} (B_i)^{\a'}_{~ \b'}
    }
    which are bifundamentals.
    We now would like to write the superpotential in terms of this field.

    One can now combine the fields into $E_I = ( A_i, \hat B_j)$, which are all
    chiral superfields in the $\bf(2,2)$ representation. We
    write the $SU(4)$ invariant expression
    \eqn\sufourinv{
   W_{SU(4)}= \epsilon_{IJKL} ( E_I)_{\a_1}^{~\b_1} ( E_J)_{\a_2}^{~\b_2} ( E_K)_{\a_3}^{~\b_3} ( E_L)_{\a_4}^{~\b_4}
     \epsilon^{\a_1 \a_4} \epsilon_{\b_1 \b_2} \epsilon^{\a_2 \a_3} \epsilon_{\b_3 \b_4 }.
     }
     This is essentially the only way to contract the gauge indices.
      In fact it is simpler to view
      the gauge group $SU(2) \times SU(2)$ as $SO(4)$ and think of
     the bifundamentals as carrying an $SO(4)$ index $m=1,2,3,4$.
     Then it is clear that the superpotential
     \eqn\defofinv{
     W_{SU(4)} =   \epsilon_{IJKL}  \epsilon_{mnrs} E_I^m E_J^n E_K^r E_L^s \sim \det( E)
     }
     is gauge-invariant and $SU(4)$-invariant.

     It remains to show that this is equivalent to the original superpotential \nwnsix .
     This can be shown as follows.
     The superpotential comes from integrating out the adjoints. In $SO(4)$ notation
     we can label the adjoint as a self dual tensor $\varphi^+_{mn}$ and $\tilde \varphi^-_{mn}$.

     We then have the superpotential couplings
     \eqn\couplfin{\eqalign{
    W =& {k\over {4\pi}}  \varphi^+_{mn} \varphi^+_{mn}
 -   {k\over {4\pi}} \tilde\varphi^-_{mn} \tilde\varphi^-_{mn}  + \varphi^+_{mn} v_{mn}  +
    \tilde \varphi^-_{mn} v_{mn}~,
    \cr
    &~~~~~v_{mn} \equiv ( A_1^m B_1^n + A_2^m B_2^n).
  }}
  After we integrate out $\varphi$ and $\tilde \varphi$ we get (up to a constant)
  \eqn\finexp{
  W = P^+_{mn, rs} v_{mn} v_{rs} - P^-_{mn,rs} v_{mn} v_{rs} \sim \epsilon_{mnrs} v_{mn} v_{rs},
  }
  where
  % Removed unnecessary redefinition
  %$v_{mn} \equiv ( A_1^m B_1^n + A_2^m B_2^n) $ and
  $P^\pm_{mn,rs}$ are projectors onto
  self dual and anti-self dual parts. This then gives
  \eqn\finexp{
  W = \epsilon_{mnrs} A_1^m B_1^n A_2^r B_2^s  \sim
 % \epsilon_{IJKL}  \epsilon_{mnrs} E_I^m E_J^n E_K^r E_L^s
 \det(E)= W_{SU(4)}.
 }
Thus we conclude that the theory has an additional $SU(4)$ symmetry. Together with the
$SU(2)_R$ symmetry that we had originally (${\cal N}=3$) this leads to a full $SO(8)$, since
the $SU(2)_R$ mixes the fields $E_I^m$ and their complex conjugates.
The $SU(4)$ symmetry of the superpotential should not be confused with the $SU(4)_R$ symmetry
of the bosonic potential that we had in \claimfin. In fact, the final bosonic potential in this
theory has an $SO(8)_R$ symmetry, and the action is the same as in \baggerlambert.

\appendix{B}{ Analysis of the M-theory geometry }

Let us begin by recalling some facts about Kaluza-Klein monopoles,
or Gibbons-Hawking metrics \gibbonshawking.
 A Kaluza-Klein monopole is specified by a circle that shrinks at its core, and three
 spatial dimensions transverse to the monopole.
 It is of the form \gibbonshawking
 \eqn\kkmon{ \eqalign{
 ds^2 = &  U d \vec x^2 + U^{-1} ( d \varphi + \vec \omega
  \cdot d \vec x  )^2 ~,~~~~~~~~ \varphi \simeq \varphi + 2 \pi,
 \cr
  & \vec \nabla^2 U \equiv \partial_a \partial_a U =0 ~,~~~~~
  \partial_{a }w^{b } - \partial_b \omega^a = \epsilon_{abc}  \partial_c U,
 }}
  where both the Laplacian and  the epsilon symbol
   are taken with respect to the flat metric on $\bf{R}^3$. The components
  of $\vec x$ have been denoted as $x_a$, so that the indices $a,b,c $ run over three values.
  These metrics describe BPS solutions. A particular solution for the function $U$
  is
  \eqn\kkams{
  U = 1 + { \kappa   \over 2  |\vec x | },
  }
  where $\kappa $ needs to be an integer in order to avoid singularities. In other words, the
  quantization condition on the flux of the ``gauge field'' associated with
  $\vec\omega$ in \kkmon\ fixes $\kappa$ to
  be an integer.
  When we go to the region $|\vec x| \to 0$ we can neglect the 1 in $U$ and set
  $U = { \kappa \over 2 |\vec x| }$.  In this case we get the metric of
  $\bf{R}^4/\bf{Z}_{ \kappa }$ or $\bf{C}^2/\bf{Z}_\kappa$, where the $\bf{Z}_\kappa$ acts as
   $(z_1,z_2) \simeq e^{ i { 2 \pi \over \kappa } } (z_1,z_2)$ on
  the coordinates of ${\bf C}^2$. We can also change the constant 1 in \kkams\ to any other constant.
  This changes the asymptotic size of the circle.

  Let us now return to our problem, described in section 3.2.
  We need to consider eight dimensional spaces $X_8$ that
  preserve $3/16$ of the supersymmetry.
  They are called  ``Toric HyperK\"ahler manifolds'', and they were studied in  \kkmonopole.
  These metrics involve two circles $\varphi_1, \varphi_2 $ (with period $2\pi$)
  and two sets of 3 spatial directions $\vec x^1 $, $\vec x^2$. The general form of the metric is
  very similar to the Kaluza-Klein monopole \kkmon, except that now the harmonic function $U$ is
  replaced by a two by two symmetric matrix of  functions $U_{ij}$ ($i,j=1,2$) \kkmonopole
  \eqn\formofm{\eqalign{
  ds^2 = & U_{ij} d \vec x^i \cdot d \vec x^j + U^{ij} ( d \varphi_i + A_i ) ( d \varphi_j  + A_j)~,
 \cr
 A_i = & d \vec x^j \cdot \vec \omega_{ji }  = dx^j_a \omega^a_{ji} ~,~~~
 ~~ ~~~ \partial_{ x_a^j} \omega^{b}_{k i } - \partial_{x_b^k} \omega^a_{ji} =
  \epsilon^{ab c} \partial_{x^j_c} U_{ki}~,
 }}
 where $U^{ij}$ is the inverse of the matrix $U_{ij}$. The matrix $U$ obeys linear
  equations that follow
 from this ansatz, see \kkmonopole .
 The metric of a single Kaluza-Klein monopole times
 % OA - minor correction
 $\bf{R}^3\times S^1$
 %$\bf{R}^4 $
 can be written in this form as a configuration with
 \eqn\onekk{
 U = U_{\infty} +   \pmatrix{ h_1 & 0 \cr 0 & 0 } ~,~~~~~~ h_1= { 1 \over 2 |\vec x_1|} ~,~~~~~~~
 U_\infty = {\bf 1} = \pmatrix{ 1 & 0 \cr 0 & 1 }.
 }
 The matrix $U_{\infty}$ encodes the shape of the torus at infinity. We can read off the value of
 the $\tau= \tau_1 + i \tau_2$
 specifying the shape of the torus by writing the asymptotic form of the metric
 \eqn\asymptf{
% OA - corrected typo in this equation
 U^{-1}_{ij} d\varphi_i d\varphi_ j = ({ \rm const })\left[ { ( d\varphi_2 + \tau_1 d\varphi_1)^2 \over
 \tau_2 }+
 \tau_2 ( d\varphi_1)^2 \right]
 }
 or
 \eqn\valtaum{
 U^{-1}_\infty =
 ( {\rm const} ) { 1 \over \tau_2}
 \pmatrix{ \tau_2^2 + \tau_1^2 & \tau_1 \cr \tau_1 & 1 } ~,~~~~~~ U_{\infty} = { 1 \over ({ \rm const}) }
 { 1\over \tau_2 }  \pmatrix{ 1 &  - \tau_1 \cr - \tau_1 & \tau_1^2 + \tau_2^2 }.
 }
 The constant is related to the area of the torus and to the size of the $6$th circle in the type
 IIB picture.
Thus, the choice in \onekk\ amounts to $\tau_1 = \chi = 0$ and
$\tau_2 = 1/g_s =1$.

 Another  very simple configuration is given by a diagonal matrix $U =
 {\rm diag}( U_1,U_2)$ with
  $U_1 = 1 + { 1 \over 2 |\vec x_1| }$ and $U_2 = { 1 + { 1 \over 2 |\vec x_2| }}$. This
 describes   two orthogonal Kaluza-Klein monopoles, each involving a different circle of the two-torus.
 The solution describing the rotated Kaluza-Klein monopole corresponding to the $(1,k)$ fivebrane
 has the form
 \eqn\rotkk{
 U= {\bf 1} +  \pmatrix { h_2  & k h_2 \cr k h_2 & k^2 h_2 },   ~~~~~
  h_2 = { 1 \over  2 |\vec x_1 + k \vec x_2 | }~.
  }
We will argue below that this metric is non-singular. Since the
equations for the matrix $U$ and for $\omega$ are linear, we can
find solutions by a superposition principle. We thus conclude that
the metric corresponding to the type IIB configuration of
 $(1,0)$ and $(1,k)$ fivebranes at an angle
 that we described in section 3 is
\eqn\twokk{
 U= {\bf 1} +   \pmatrix{ h_1 & 0 \cr 0 & 0 } +
 \pmatrix { h_2  & k h_2 \cr k h_2 & k^2 h_2 },   ~~~~~
  h_2 = { 1 \over  2 |\vec x_1 + k \vec x_2 | } ~,~~~~h_1 = { 1 \over 2 | \vec x_1 | }~.
  }
  We can change the asymptotic values of the radii of the circles by  changing ${\bf 1}$ to
  a more general constant matrix $U_{\infty}$.
Note that the form of the metric and the equations for $U$ are
such that they
 are invariant under $GL(2)$ transformations of coordinates such that the upper and lower indices $i,j$ transform
 covariantly or contravariantly. (The positions of the $i,j$ indices
 % OA - minor correction
 are
 %is
 important, but
 the positions of the $a,b$ indices in \formofm\ are not.)
 Such a transformation applied to the above solution
  would change the asymptotic shape of the $\varphi_i$
 two-torus into a more general one, and it
 corresponds to the usual $GL(2) = {\bf R} \times SL(2)$ moduli space of supergravity on a $T^2$.
 For the time being we will fix the asymptotic values as
 in  \twokk.
 It is however important to note that if we perform transformations that are in
 $SL(2,{\bf Z})$ on \onekk, we can get to
 \rotkk\ without changing the asymptotic shape of the torus,
 since the redefinition of the coordinates
 $\varphi_1, \varphi_2$ is consistent with their periodicities.
 Since the original metric \onekk\ is non-singular, it
 is clear that \rotkk\ is also non-singular.

 Now let us consider the metric \twokk\ which is the superposition of both KK monopoles.
 We see that near each of the monopoles we reduce to one of the previous solutions
 (with $\kappa=1$), and the metric is
 not singular. So, any singularity could only arise at the point $\vec x_1 \sim \vec x_2 \sim 0$ where the monopoles
 intersect. When we are near this point we can neglect the ${\bf 1}$ contribution in \twokk .
 We can now do a $GL(2)$ transformation of coordinates of the form
 \eqn\matrg{
 \vec{x}'^1 = \vec{x}^1~, ~~~~~\vec{x}'^2 = \vec{x}^1 + k \vec{x}^2~.
 }
 With this choice we see that
 \eqn\finalfo{
  U' = \half \pmatrix{ { 1 \over | \vec x'^1| } & 0 \cr 0 & { 1 \over |\vec x'^2| } }.
  }
  This $U'$ matrix gives us the superposition of two orthogonal KK monopoles in completely
  orthogonal directions in the ``near horizon limit'', where they seem to be simply $\bf{R}^4 \times \bf{R}^4$.
  However, this is not completely true, because this $GL(2)$ transformation does not preserve
  the periodicities of the $\varphi_i$'s. The new $\varphi'$ coordinates, which
  are determined by the inverse transformation
  \eqn\explc{
  \varphi'_1 = \varphi_1 - { 1 \over k } \varphi_2 ~,~~~~~~~\varphi'_2 = { 1 \over k } \varphi_2~,
  }
 do not have the same periodicity conditions as the original ones.
 The new identifications can be deduced from
 % OA - removed extra "the"
 %the
the original identifications for $\varphi_1$ and $\varphi_2$.
  From this we conclude that $\varphi'_1$ and $\varphi'_2$ have the usual identifications by $2\pi$ that would have
  given a transverse space $\bf{R}^8$,
  plus the extra identification
  \eqn\extraident{
  ( \varphi'_1 ,
  \varphi'_2 ) \sim ( \varphi'_1, \varphi'_2 ) + 2 \pi ( - { 1 \over k } , { 1 \over k } )~.
  }

  We conclude that the KK monopole configuration that is U-dual to the
  brane configurations that we discussed has a $\bf{C}^4/\bf{Z}_k$ singularity.
  In particular, if we define four
  complex coordinates $z_I$ parameterizing $\bf{C}^4$, we see that the identification
  acts as $z_I \to e^{ 2 \pi i/k} z_I$. In the special case of $k=1$ the manifold
is completely non-singular, as it looks like
 $\bf{R}^8$ at the origin.

Note that if we had included the ${\bf 1}$ in \twokk\ then this
would have lead to an additional constant contribution to
\finalfo\ of the form $U'_{\infty} = \pmatrix{ 1 + { 1 \over k^2}
& - { 1 \over k^2 } \cr - { 1 \over k^2}  & { 1\over k^2} } $. As
explained in \kkmonopole, a configuration of two orthogonal KK
monopoles as in \finalfo\ plus a constant matrix $U'_\infty$
preserves ${\cal N}=3$ (or 3/16) supersymmetry if the matrix
$U'_{\infty}$ is not diagonal, and ${\cal N}=4$ (or 1/4)
supersymmetry if the matrix is diagonal.

The ${\cal N} = 4$ brane configuration of subsection 3.5 is very
similar except that after the transformation \matrg\ the value of
$U'_\infty$ is diagonal, of the form $U'_\infty = \pmatrix{ a & 0
\cr 0 & 1/a } $. As remarked above the configuration is then
${\cal N}=4$ supersymmetric ($1/4$ of the supercharges are preserved). In that case, the full
geometry, including the asymptotics, is given by the $\bf{Z}_k$
quotient of an orthogonal product of Taub-NUTs, as $U'$ is
diagonal.
 This means that the original  configuration is given
by \twokk\ with ${\bf 1}$ replaced by $U_\infty$ with \eqn\newrep{
U_\infty = \pmatrix{a + { 1 \over a}  & { k \over a }  \cr { k
\over a }  & { k^2 \over a }  }. } Using \valtaum\ we see that this
corresponds to a $\tau$ parameter of the form \eqn\tauparm{ \tau =
\chi + { i \over g_s } = { - k + i k a \over 1 + a^2 }~. } Thus, the
only difference between the ${\cal N}=3$ and the ${\cal N}=4$
configurations lies in the asymptotic values of the parameters of
the two-torus, or the constant matrix $U_{\infty}$ that replaces
the ${\bf 1}$ in \twokk .
 The geometry around $ \vec x_i \sim 0$ is exactly the same for both
configurations. Note that with the matrix \newrep , the asymptotic
form of the metric in the two ${\bf R^3}$ spaces is not diagonal
in the $\vec x^i$ coordinates.  We can diagonalize the metric by
doing again the change of coordinates \matrg , but now only on the
coordinates $\vec x^i$ and not the angles. This would then give
the picture used in subsection 3.5, with a $(1,0)$ brane and a $(1,k)$
brane which are orthogonal in the two ${ \bf R^3}$ spaces
  but in a background
which has a non-trivial RR axion.

More generally, for an asymptotic metric with $U_\infty$ as in
\valtaum\ we see that the asymptotic metric in the ${\bf R^3
\times R^3 }$ subspace from \formofm\ is not diagonal. We can
diagonalize it by choosing new coordinates of the form \eqn\newco{
\tilde { \vec x}_1 = \vec x_1 - \tau_1 {\vec x}_2 ~,~~~~~~\tilde
{\vec x}_2 = \tau_2 {\vec x}_2~. } This means that the
% OA - minor change
combinations
%combination
of coordinates that appear in the harmonic
% OA - minor change
functions
%function of
$h_1$ and
$h_2$ corresponding
% OA - minor addition
to
the $(1,0)$ and $(1,k)$ fivebranes
% OA - minor change
are
%is
\eqn\comfinc{ h_1 \sim { 1 \over | \tau_2 \tilde {\vec x}_1 +
\tau_1 \tilde {\vec x}_2 | } ~, ~~~~~~~~~~~ h_2 \sim { 1 \over
|\tau_2 \tilde {\vec x}_1 + ( \tau_1 + k ) \tilde {\vec x}_2 |}~, }
where we neglected an overall constant. This implies that the
angle between the $(1,0)$ and the $(1,k)$ fivebranes is given by
\eqn\anglefi{ \theta= {\rm arg}(\tau) - {\rm arg}(\tau  + k )~. }
For \tauparm\ we get that $\theta = \pi/2$ so that we have
orthogonal branes.

\listrefs

\bye